\documentclass[12pt,preprint]{aastex}

\usepackage{lineno}

\newcommand{\be}{\begin{equation}}
\newcommand{\ee}{\end{equation}}
\def\lta{\,\raise 0.3 ex\hbox{$ < $}\kern -0.75 em
 \lower 0.7 ex\hbox{$\sim$}\,}
\def\gta{\,\raise 0.3 ex\hbox{$ > $}\kern -0.75 em
 \lower 0.7 ex\hbox{$\sim$}\,} 
\newcommand{\mplan}{ M_{\rm p}} 
\newcommand{\bplan}{ B_{\rm p}} 
\newcommand{\rplan}{ R_{\rm p}} 
\newcommand{\tplan}{ T_{\rm p}} 
\newcommand{\dotplan}{ {\dot M}_{\rm p}}
\newcommand{\dotzero}{ {\dot M}_0}
 
\newcommand{\mdisk}{ M_{\rm d}} 
\newcommand{\rdisk}{ R_{\rm d}} 
 
\newcommand{\vsound}{ v_{\rm s}} 

\newcommand{\ups}{u_{\rm x}}
\newcommand{\power}{{\cal P}} 
\newcommand{\zetarm}{\zeta_{\scriptscriptstyle R}} 
\newcommand{\bkap}{b_\kappa} 
\newcommand{\kapindex}{\eta} 
\newcommand{\fpc}{f_{\rm pc}}

\begin{document} 

\title{Analytic Approach to the Late Stages of Giant Planet Formation}

\author{Fred C. Adams$^{1,2}$ and Konstantin Batygin$^3$}  
\affil{$^1$Physics Department, University of Michigan, Ann Arbor, MI 48109} 
\affil{$^2$Astronomy Department, University of Michigan, Ann Arbor, MI 48109} 
\affil{$^3$Division of Geological and Planetary Sciences\\
California Institute of Technology, Pasadena, CA 91125} 
\affil{$\,$} 

\begin{abstract}

This paper constructs an analytic description for the late stages of
giant planet formation. During this phase of evolution, the planet
gains the majority of its final mass through gas accretion at a rapid
rate. This work determines the density and velocity fields for
material falling onto the central planet and its circumplanetary disk,
and finds the corrsponding column density of this infalling envelope.
We derive a steady-state solution for the surface density of the disk
as a function of its viscosity (including the limiting case where no
disk accretion occurs). Planetary magnetic fields truncate the inner
edge of the disk and determine the boundary conditions for mass
accretion onto the planet from both direct infall and from the
disk. The properties of the forming planet and its circumplanetary
disk are determined, including the luminosity contributions from
infall onto the planet and disk surfaces, and from disk viscosity.
The radiative signature of the planet formation process is explored
using a quasi-spherical treatment of the emergent spectral energy
distributions. The analytic solutions developed herein show how 
the protoplanet properties (envelope density distribution, velocity 
field, column density, disk surface density, luminosity, and 
radiative signatures) vary with input parameters (instantaneous 
mass, orbital location, accretion rate, and planetary magnetic 
field strength). 

\end{abstract} 

$\,\,\,$ {\it UAT Concepts:} Exoplanet formation (492); 
Exoplanet astronomy (486)

\section{Introduction} 
\label{sec:intro} 

Our understanding of the planet formation process remains incomplete.
The current paradigm for giant planet formation breaks the process
into separate stages (see, e.g., \citealt{pollack1996},
\citealt{benz2014}, and references therein). In the first stage, a
high metallicity core is constructed within the circumstellar disk.
The growing body retains a rock/ice composition until it reaches a
mass threshold of approximately $\mplan\sim10M_\oplus$, when it enters
the second stage and starts to capture a gaseous atmosphere (made
primarily of hydrogen and helium). In this stage, the gaseous envelope
displays an extended spatial distribution, and must cool and contract
in order for additional mass to accrete. Once the envelope reaches a
mass comparable to the initial rocky core, so that the object has
total mass $\mplan\gta20M_\oplus$, the self-gravity of the envelope
facilitates rapid contraction and allows for runaway accretion of
material from the background disk. Most of the mass of the giant
planet is accreted during this third phase, which is the subject of
this present work.

More specifically, the goal of this paper is to construct an analytic
model that describes the collapse of material onto the planet after
the flow detaches from the background circumstellar disk and enters
the planetary sphere of influence. This work thus focuses on the range
of size scales roughly defined by 
\be
\rplan \le r \le R_H \,.
\label{radrange} 
\ee 
The planetary surface at radius $\rplan$ provides the inner boundary
and the Hill radius $R_H$ provides an effective outer boundary. The
background circumstellar disk (beyond $R_H$) exhibits complex
hydrodynamic motion that ultimately funnels gaseous material into the
Hill sphere at some rate. In practice, however, not all of the
material that enters the Hill sphere is accreted, as some of the
material flows back out.  The accretion rate of material that remains
within the Hill sphere is denoted here as $\dotplan$. The resulting
inward flowing material then falls toward the central planet and its
accompanying circumplanetary disk. The inner boundary, given by the
planetary radius $\rplan$, varies only weakly with planet mass and is
determined by the physics of planetary structure.

Working within the regime defined above, this paper first presents
analytic solutions for the density distribution $\rho(r,\theta)$, the
velocity field ${\vec v}(r,\theta)$, and the corresponding column
density $N_{\rm col}(\theta)$ of the infalling envelope (for a
spherical coordinate system centered on the planet and under the
assumption of azimuthal symmetry). We then find a steady state
solution for the surface density $\Sigma(r)$ of the circumplanetary
disk along with the radial scales of the problem. The inner boundary
condition for accretion from the disk onto the planet is determined by
the magnetic truncation radius $R_X$, whereas direct accretion is
controlled by the magnetic capture radius $R_\sigma$. The outer disk
radius $\rdisk$ is specified by the centrifugal barrier $R_C$ of the
collapse flow.  All of the aforementioned quantities collectively 
determine the luminosity contributions of the forming planetary
object, due to infall and accretion, for both the disk and the
planet. These results are then combined to estimate the emergent
spectral energy distributions of the forming planet. The resulting
analytic solutions are functions of four input parameters, including
the planet mass $\mplan$, the accretion rate $\dotplan$, the planetary
magnetic field strength $\bplan$, and the semimajor axis $a$ of the
forming planet. In addition to determining how protoplanetary
properties depend on the input parameters, these solutions can be used
to explore evolutionary scenarios (e.g., accretion rate as a function
of planet mass), in a variety of physical regimes and with negligible
computational cost.

A great deal of previous work has been carried out concerning the
problem of giant planet formation. This present paper focuses on the
third stage of the process --- when the majority of the mass is
accumulated --- with the starting conditions consistent with those
found for the earlier stages \citep{pollack1996,benz2014}. Much of the
previous work concerning the rapid accretion phase has been numerical
(e.g., \citealt{hubickyj2005,lissauer2009,szulagyi2016,lambrechts2019}).  
In contrast, this work adopts an analytic approach and is thus
complementary to previous numerical efforts. 

The solutions of this paper are applicable in the regime where
the material falling toward the planets approaches pressure-free
conditions, so that gas parcels follow nearly ballistic
trajectories. This approximation holds in the limit where the gas can
cool sufficiently (e.g., under isothermal conditions).  Due to
conservation of angular momentum, the collapse produces a well-defines
centrifugal radius $R_C$, which defines the nominal radius of the
circumplanetary disk. The disk supports an accretion flow for inner
radii $r<R_C$ (due to viscosity) and spreads outward for larger radii.
The present treatment does not include torques acting on the gas
parcels during infall, so that the resulting values of $R_C$
correspond to upper limits. The solutions are robust, however, in that
other values of the centrifugal radius $R_C$, equivalently other
starting profiles for the pre-collapse angular momentum, can be
accommodated.

This paper is organized as follows. We briefly outline (in Section
\ref{sec:csdisk}) the properties of the background circumstellar disk
that provides the environment for the late stages of gas accretion
onto planets. In Section \ref{sec:infall}, we construct a set of
analytic solutions that describe the infall of material that falls
onto the planet. Angular momentum considerations demand that the
majority of incoming material falls first to form an accompanying
circumplanetary disk, whose properties and evolution are considered in
Section \ref{sec:cpdisk}. The forming planet/disk system generates
significant luminosity via accretion; the various contributions to the
total power and the corresponding radiative signature of forming
planets are determined in Section \ref{sec:radiation}. The paper
concludes, in Section \ref{sec:conclude}, with a summary of our
results, a discussion of their implications, and a comparison between
the planet/disk systems that arise during planet formation and the
star/disk systems that arise during star formation. Given that
approximations must be made in order to obtain analytic results,
Appendix \ref{sec:control} provides quantitative estimates for the
accuracy of this approach and defines its regime of validity.
Magnetic fields play an important role in controlling gas flow near
the planet, and the magnetic field strengths expected on young planets
are estimated in Appendix \ref{sec:magfield}. Finally, we present 
generalizations of the standard assumptions in Appendices 
\ref{sec:hillpressure} through \ref{sec:evolution}. 

\section{Circumstellar Disk Environment} 
\label{sec:csdisk} 

Consider a planet that has grown past the core formation and initial
envelope cooling stages so that it is actively gaining gaseous
material from the background disk environment. The mass flow from the
circumstellar disk onto the planet can be conceptually divided into
the following parts. The circumstellar disk supports an inward
accretion flow (through the disk and eventually onto the star) at a
well-defined rate. As this large scale accretion flow crosses the
radial location of the forming planet, some fraction of the material
enters the sphere of influence of the planet. As a first
approximation, we consider the boundary between the disk and the
planet to be given by the Hill radius $R_H$, 
\be 
R_H = \left( {\mplan \over 3M_\ast}\right)^{1/3} a \, ,
\label{rhill} 
\ee
where $a$ is the semimajor axis of the planetary orbit, $\mplan$
is the mass of the planet, and $M_\ast$ is the mass of the star.

Not all of the material that enters into the Hill radius will accrete
onto the planet. First, some of the gas that enters the Hill sphere
will promptly flow back out to the background disk and never become
part of the sphere of influence of the planet \citep{lambrechts2017}. 
As shown below (Section \ref{sec:infall}), conservation of angular
momentum dictates that most of the material will fall to radial
locations much larger than the radius of the forming planet and must
accrete through a circumplanetary disk.  This suppression of direct
accretion onto the planet \citep{machida2008}, along with ineffiencies
in the disk accretion process \citep{szulagyi2017,fung2019}, act to
reduce the amount of mass received by the planet. In addition, the
planet can produce a gap in the circumstellar disk with annular width
given by the planet mass, disk viscosity, and other parameters. With
its lower surface density, the the gap reduces the density of material
flowing through the Hill boundary into the vicinity of the planet
\citep{malik2015}. Finally, magnetic fields anchored within the planet 
can suppress accretion onto the planetary surface
(\citealt{batygin2018,cridland2018}; see Section \ref{sec:magradius}).

All of the above effects conspire to produce a net mass accretion rate
$\dotplan$ that enters the Hill sphere of the growing planet. For a
given rate $\dotplan$ and given instananeous mass $\mplan$ of the
planet, we can solve for the rotating inward flow onto the planet and
disk, as well as the properties of the disk (see the following
sections). A separate but related issue is to understand how the mass
accretion rate $\dotplan$ depends on planetary mass $\mplan$. The
geometric cross section for the planet to capture material from the
circumstellar disk scales as $R_H^2$ (e.g., \citealt{zhu2011}).  In
addition, the gas flowing inward through the boundary defined at the
Hill radius can shock, and thereby increase its density by a factor of
$(v/\vsound)^2$, where $\vsound$ is the sound speed and $v\sim\Omega$
$R_H$ is the impact speed at $R_H$ (where $\Omega$ is the mean motion). 
As a result, the mass accretion rate entering the Hill sphere is
expected to scale as $R_H^4$ in the presence of shocks
\citep{tanwat2002,tantan2016,lee2019}.  In this present application,
we can independently specify the instantaneous mass accretion rate and
planet mass in order to calculate the dynamics of the infall, disk,
and luminosity generation. With the resulting solutions in hand, one
can choose a particular scenario for $\dotplan$ as a function of
$\mplan$ to consider evolutionary sequences (see Appendix
\ref{sec:evolution}).

For completeness, we note two additional length scales of interest.
The first is the scale height $H$ of the disk, which is given by  
\be
H = \left({a^3 \over GM_\ast}\right)^{1/2} \vsound \,, 
\label{height} 
\ee
where $a$ is the semimajor axis of the growing planet and $\vsound$ is
the sound speed.  In general, disk accretion (in the circumstellar
disk) is thought to occur within the surface layers of the disk, where
ionization is sufficient to support magentically generated turbulence
and/or where magnetically driven winds operate. This constraint limits 
disk accretion to the outer surface density increment of
$\Sigma_0\approx$ 100 g cm$^{-2}$, which defines a second length scale
of interest --- the thickness $T$ of the active layer.  The surface
density of the Minimum Mass Solar Nebula at $r=5$ AU generally falls
in the range $\Sigma_5\approx180-350$ g cm$^{-2}$. As a result, the
total active layer $2T$ can be an appreciable fraction the disk scale 
height, i.e.,
\be 
2T = {2\Sigma_0 \over \Sigma_5} H \approx (0.5 - 1)H\,.
\label{thickness} 
\ee
For the late stages of giant planet formation at $a\sim5$ AU, 
the length scales obey the ordering $R_H\sim{H}\sim{2T}$.  
Note, however, that the thickness of ionization layer due 
to UV radiation (in contrast to cosmic rays) can be thinner
\citep{perez2011}. For planets forming in smaller orbits ($a\sim0.05$
AU), local ionization allows for the disk to be MRI active and the
thickness $T$ is no longer relevant.  Since $R_H\propto{a}$ and
$H\propto{a^{3/2}}$, $R_H>H$ for sufficiently tight orbits.

The background disk environment also constrains the time scale for
giant planet formation. Observations show that disks retain
substantial amounts of gas, and thus allow for gas accretion, over
time scales in the range $1-10$ Myr \citep{haisch2001,jesus}. The
observations are consistent with disks having an exponential
distribution of lifetimes, $dF/dt\sim\exp[-t/\tau]$, with time scale
$\tau\approx5$ Myr (with a corresponding `half-life' $t_{1/2}$ =
$\tau\ln2\sim3.5$ Myr). As a result, the typical mass accretion rate
onto the planet must be of order $\dotplan\sim1M_J$/Myr.  If the mass
accretion rate was much smaller, then the circumstellar disk would run
out of gas before Jovian mass planets could be made.  The mass
accretion rate could in principle be somewhat larger, but if the
planet formation time is much shorter than the disk lifetime, the
issue of how and when accretion ends becomes problematic. The
solutions presented here are applicable over a wide range of
accretion rates, but we focus the discussion on values comparable 
to the above benchmark. 

\section{Infall Collapse Solution}  
\label{sec:infall} 

This section considers a planetary core that is actively gaining
material from the background circumstellar disk. We assume that the
forming planet has entered the third phase, with rapid gas accretion,
so that $\mplan\gta20M_\oplus$.  Starting with infall-collapse
solutions obtained previously for the star formation problem
\citep{ulrich1976,cassen1981,terebey1984}, we can construct the
density and velocity fields for the inward flow.

As it enters the sphere of influence of the growing planet, the
initial (pre-collapse) gas rotates with angular velocity $\Omega$,
which is determined by the mean motion of the planetary orbit. As the
material falls toward the planet, within its sphere of influence, the
gas must conserve its specific angular momentum. As a result, not all
of the incoming material will directly reach the planetary surface.
Instead, it must collect into a circumplanetary disk, and then work
its way onto the planetary surface through some type of accretion
mechanism.

As a first approximation, we neglect pressure forces and consider
parcels of gas that enter the Hill sphere to fall toward the planet on
zero energy ballistic orbits (in the rest frame of the planet). The
gas must conserve its angular momentum and hence orbit in a plane
defined by its initial polar angle $\theta_0$ or equivalently
$\mu_0=\cos\theta_0$. As the gas orbits within its plane, which is
tilted with respect to the coordinate system centered on the planet,
it follows an orbit equation which can be written in the form 
\be
1 - {\mu \over \mu_0} = \zeta \left( 1 - \mu_0^2\right) \qquad 
{\rm with} \qquad \zeta \equiv {j_\infty^2 \over G \mplan r} \,,
\label{orbit} 
\ee
where $\mu_0$ specifies the starting angle, $j_\infty$ is an angular
momentum scale, and the instantaneous location of the parcel is given
by $(r,\mu)$. Since the incoming flow detaches itself from the
circumstellar disk at radius $R_H$, the maximum specific angular
momentum is given by 
\be
j_\infty = \Omega R_H^2 \,,
\label{jinfinity} 
\ee
where this maximum value occurs for gas parcels starting in the
equatorial plane. For any starting condition, the incoming parcels
intersect the circumplanetary disk plane when $\mu=0$.  The parcels
with the largest specific angular momentum will cross the disk plane
at the disk radius, equivalently the centrifugal radius, which is
given by 
\be
R_C = {j_\infty^2 \over G \mplan} = 
{\Omega^2 R_H^4 \over G \mplan} = {R_H \over 3} \,.
\label{rcent} 
\ee
As the planet grows in mass, the Hill radius $R_H$ grows, and the
circumplanetary disk grows such that its outer radius is always given
by $\rdisk=R_C=R_H/3$ \citep{quillen1998,martin2011}.

Note that this treatment implicitly assumes that infall takes place
over the entire $\Omega_S=4\pi$ solid angle centered on the planet. If
infall is restricted to the polar regions, defined for example by the
range of starting polar angles $1\ge{\mu_0}\ge\mu_b$, then the disk
radius becomes $\rdisk$ = $(1-\mu_b^2)R_C$ = $(1-\mu_b^2)R_H/3$. This
treatment also assumes that the orbits are ballistic (pressure-free).
This approximation is quantified in Appendix \ref{sec:control}.
Moreover, including corrections for non-zero pressure and/or magnetic
fields can be incorporated into the approximation scheme (see 
Appendix \ref{sec:hillpressure}). Pressure effects counter the
effective gravitational field of the planet and thereby reduce the
range of influence to $R<R_H$. Magnetic fields can also induce
braking effects (angular momentum transfer) and reduce the effective
value of $\Omega$ in equation (\ref{jinfinity}). All of these
corrections thus reduce the disk centrifual radius $R_C$. The analytic
results of this paper are valid for any choice of $R_C$.  However, for
the sake of definiteness, we use the nominal value from equation
(\ref{rcent}) as a first approximation.

Using the orbit equation (\ref{orbit}) and conservation of energy, 
we can solve for the velocity fields of the collapse flow. If we 
first define the velocity scale
\be
v_0 \equiv (G\mplan/r)^{1/2} \,,
\label{vzero} 
\ee
then the velocity components can be written in the form 
\be
v_r = - v_0 \left[ 2 - \zeta (1-\mu_0^2) \right]^{1/2} \,,
\label{vradial} 
\ee
\be
v_\theta = v_0 \left[ {1-\mu_0^2 \over 1-\mu^2}
(\mu_0^2-\mu^2) \zeta \right]^{1/2} \,,
\label{vtheta} 
\ee
and
\be
v_\phi = v_0 (1-\mu_0^2) (1-\mu^2)^{-1/2} \zeta^{1/2} \,.
\label{vphi} 
\ee
Note that the variables $\zeta$, $\mu$, and $\mu_0$ are defined
through the orbit equation (\ref{orbit}), so that the velocity field
is specified completely (but implicitly) for any location $(r,\mu)$ in
the envelope. The minus sign in front of the radial velocity $v_r$ 
indicates that the flow is radially inward. 

The density of the infalling material, surrounding the growing 
planet, is given by conservation of mass along streamlines
\citep{chevalier1983} and takes the form 
\be
\rho(r,\mu) = { \dotplan \over 4\pi r^2 |v_r|} 
{d\mu_0\over d\mu} = { \dotplan \over 4\pi r^2 |v_r|} 
\left[ 1 + 2\zeta P_2(\mu_0) \right]^{-1} \,,
\label{density} 
\ee
where the relation between $\mu$ and $\mu_0$ is determined by 
the orbit equation (\ref{orbit}), and where $\zeta = R_C/r = R_H/3r$. 
The function $P_2(x)$ is the Legendre polynomial of degree 2 
\citep{abrasteg}. 

\begin{figure}
\includegraphics[scale=0.70]{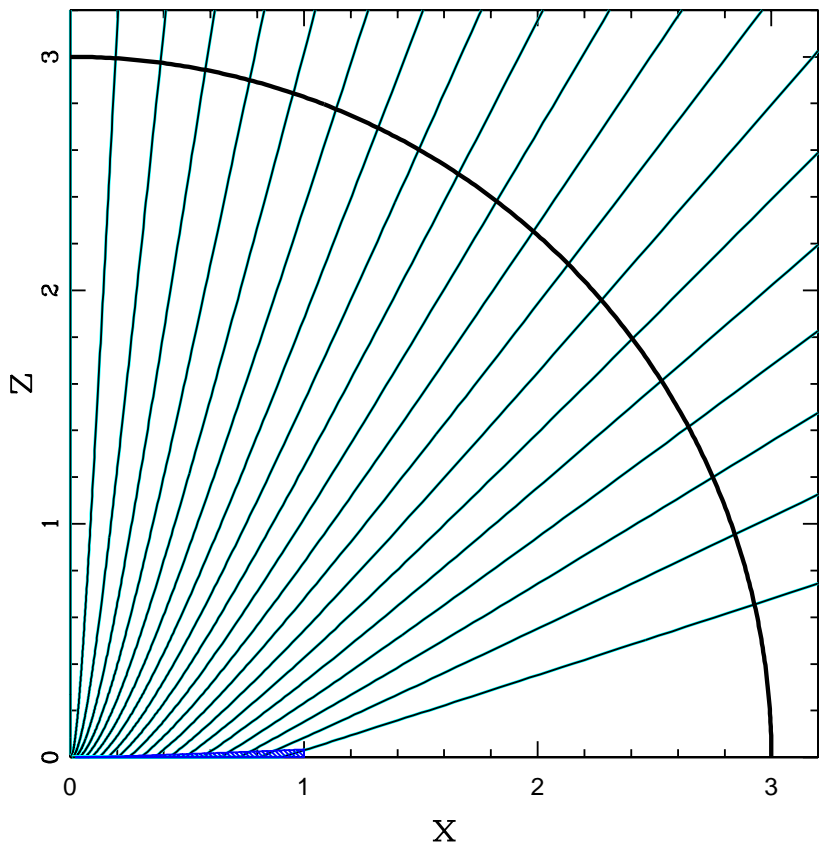}
\vskip-1.50truein
\caption{Projected trajectories of incoming parcels of gas falling 
toward the planet and its growing circumplanetary disk. The collection 
of trajectories corresponds to equally spaced initial values of 
$\sin\theta_0=0.05n$ (for integers $n$ = 0 -- 19). The length scales 
are given in units of the centrifugal radius $R_C$. The disk, with 
radius $\rdisk=R_C$, is depicted as the thin blue shaded region. 
The solid black curve at radius $r=3$ marks the Hill radius $R_H$. 
On this scale, the planet radius $\rplan\approx0.006$. } 
\label{fig:infall} 
\end{figure} 

The projected trajectories of the incoming parcels of gas are depicted
in Figure \ref{fig:infall}. The initial (outer boundary) conditions
for the trajectories are evenly spaced in $\sin\theta_0$.  Note that
the trajectories pass through the midplane ($z=0$) of the system
before reaching their periastron point. Since we assume that the
system has reflection symmetry about the $z=0$ plane, for each parcel
of gas passing through the midplane from above, there will be a
corresponding parcel passing through from below. These gas parcels
will collide and shock, and thereby form a circumplanetary disk in
the equatorial plane (delineated as the thin blue structure in the
region $0\le{x}\le1$ in the figure). Notice that the trajectories
shown in Figure \ref{fig:infall} are projected onto the meridional
plane. The full three dimensional trajectories include an azimuthal
component, so that the paths spiral around the rotational pole of the
system as they move inward.  This picture compares well with results
found previously using numerical simulations (e.g., see Figure 4 of
\citealt{tanigawa2012}).

The above analysis provides a two dimensional description of the density
distribution. It is sometimes useful to find an equivalent spherical
infall region (\citealt{adams1986}; Section \ref{sec:radiation}) by
averaging the density distribution over the polar angle $\theta$
(equivalently, over $\mu$).  The resulting density profile takes the
form 
\be
\rho = C r^{-3/2} {\cal A}(u) \,,
\label{rhosphere} 
\ee
where the constant $C$ is defined by
\be
C \equiv {\dotplan \over 4\pi \sqrt{2G\mplan}} \,,
\label{cconstant} 
\ee
and the asphericity factor ${\cal A}(u)$, with $u=r/R_C$, 
takes the form 
\be
{\cal A}(u) = \sqrt{2u} \log 
\left[ {1 + \sqrt{2u} \over D(u)} \right] \qquad {\rm where} \qquad 
D(u) = \left\{ \matrix{ \sqrt{1-u} + \sqrt{u} & {\rm for} \,\, u<1 \,; \cr
$\,$ & $\,$ \cr 
\sqrt{2u-1} & {\rm for} \,\, u>1 \,.} 
\right. 
\label{asphericity} 
\ee
Note that ${\cal A}(u)\to1$ in the limit of large radius $u\to\infty$
and that ${\cal A}(u)\to(2-\sqrt{2})u$ in the limit of small radius
$u\to0$. Notice also that this spherically averaged density
distribution is not simply the density distribution that one would
obtain from solving the spherically symmetric problem: Due to rotation
and conservation of angular momentum, material does not fall as far
inward, so that the density does not increase as rapidly in the limit
$r\to0$ (in this solution, compared to that for spherical collapse). 

The column density through the infalling envelope plays an important 
role in determining the radiative signature of the forming planet 
(see Section \ref{sec:radiation}). This column density is given by 
\be
N_{\rm col} = \int \rho dr = C R_C^{-1/2} \int u^{-3/2} {\cal A}(u) du \,.
\ee
If we take the limits of integration from 0 to $\infty$, the 
integral can be evaluated to obtain 
\be
N_{\rm col} = {\sqrt{2}\pi^2 \over 4} C R_C^{-1/2} \,. 
\label{column} 
\ee
For typical input parameters ($a=5$ AU, $\mplan=1M_J$,
$\dotplan=1M_J$/Myr), the column density $N_{\rm col}\approx0.05$ g
cm$^{-2}$.  Using a standard value of the dust opacity $\kappa_V=200$
cm$^2$ g$^{-1}$ \citep{draine1984}, the visual extinction
$A_V\approx10$. However, the surface temperature of the planet is only
$\tplan\sim1000$ K, corrsesponding to wavelengths $\lambda\sim3\mu$m,
so that the optical depth of the envelope to the planetary radiation
field is expected to be of order unity. As a result, only a modest
fraction of the planetary luminosity will be absorbed by the infalling
envelope and re-radiated at infrared wavelengths. Morever, the
infrared extinction is less than unity, so that the envelope should be
optically thin to the radiation it emits (see Section
\ref{sec:radiation}).

The column density is lowest along the rotational pole of the system,
where $N_{\rm col}(\mu=0)=\pi C R_C^{-1/2}$, which is smaller than the
equivalent spherical value by a factor of $\pi\sqrt{2}/4\sim0.9$. The
column density is larger along equatorial lines of sight, and becomes
formally infinite in the limit $\mu=0$. In practice, however, this
limit is not realized because the circumplanetary disk will retain a
finite scale height. In addition, the equatorial lines of sight are
blocked by the background circumstellar disk that surrounds the
forming planet. We can also determine the corrections to the column
density due to the finite inner cutoff at $r=R_X$ and the outer cutoff
at $r=R_H$. Using the limiting forms for ${\cal A}(u)$ from equation
(\ref{asphericity}), we find the contributions 
\be
\int_0^{R_X} \rho dr = C R_C^{-1/2} 2(2-\sqrt{2})\sqrt{\rplan/R_C} 
\qquad {\rm and} \qquad 
\int_{R_H}^\infty \rho dr = C R_C^{-1/2} 2\sqrt{3}/3\,. 
\label{colcorrect} 
\ee
The first correction is small (only $\sim2.5\%$ of the total column
density), whereas the second correction is more substantial ($\sim1/3$
of the total). As a result, one should multiply the column density of
equation (\ref{column}) by a correction factor 
$(1-4\sqrt{6}/3\pi^2)\approx0.67$.

A related quantity is the mass of the infalling envelope at a given
time in the development of the planet. The mass that is contained
within the Hill sphere but has not reached the planet or disk is given
by the integral
\be
M_{\rm env} = \int_{\rplan}^{R_H} C r^{-3/2} {\cal A}(u) 4\pi r^2dr
\approx \sqrt{6} \dotplan \Omega_C \,, 
\ee
where $\Omega_C^2=G\mplan/R_C^3$. The ratio of the envelope mass 
to the planet mass is given by the ratio of the orbital period at 
the disk edge to the total planet formation time. We thus expect 
$M_{\rm env}/\mplan\sim10^{-6}\ll1$. 

Note that the infalling envelope considered here is not in hydrostatic
equilibrium, but rather is freely falling toward the central planet
(and its disk) at the center of the flow. This configuration applies
only to the final stage of giant planet formation when gas accretes
rapidly, and when the planet gains most of its mass. In contrast, in
the earlier stages, the material in this region is quasi-hydrostatic
and must cool before it condenses toward the forming planet.

\section{Circumplanetary Disk}
\label{sec:cpdisk} 

This section outlines the formation and structure of the
circumplanetary disk. As shown here, most of the material falls onto
the disk, rather than directly onto the planet, so that disk accretion
is an important feature of the process.

\subsection{Infall onto the Disk} 
\label{sec:infalldisk} 

The rate at which the circumplanetary disk receives infalling material
from the envelope is given by the density field and the velocity field.
Using the results from the previous section, we obtain
\be
{d \Sigma \over dt} = 2 \rho v_\theta \Bigg|_{(r,\mu=0)} = 
{ \dotplan \over 2\pi r^2}{v_\theta\over |v_r|}  
{d\mu_0\over d\mu} = { \dotplan \over 4\pi r^2} 
{1 \over \zeta \mu_0} \,,
\ee
where all quantities are evaluated at the disk plane $(\mu=0)$. Note
that the infall strikes the disk from both sides (top and bottom),
leading to a factor of 2 in the result. Again using the orbit equation
(\ref{orbit}), the above expression can be evaluated to obtain 
\be
{d \Sigma \over dt} = { \dotplan \over 4\pi r^2} 
{1 \over (\zeta^2-\zeta)^{1/2}} \,.
\ee

As a benchmark, it is useful to determine the surface density of the
disk that would result in the absence of disk accretion (see the
following section). This surface density is given by the integral of
$d\Sigma/dt$ over time, equivalently mass, so that we obtain 
\be
\Sigma_0 = \int { \dotplan dt \over 4\pi r^2} 
{1 \over (\zeta^2-\zeta)^{1/2}} \,.
\ee
The subscript indicates that this surface density would result 
in the absence of viscous evolution $(\nu=0)$. 
Using $dM = \dotplan dt$ and the definition of $\zeta$, 
\be
\zeta = {R \over 3r} = {a \over 3r} 
\left({M \over 3M_\ast}\right)^{1/3} 
\qquad \Rightarrow \qquad 
M = (3r/a)^3 3M_\ast \zeta^3 \,,
\ee
we find 
\be
\Sigma_0 = {243 M_\ast r \over 4 \pi a^3} 
\int {\zeta^2 d\zeta \over (\zeta^2-\zeta)^{1/2}} \,.
\ee
Next we change variables to $u$ defined such that $u=1/\zeta$, 
so that the limits of integration are $u = 3r/R_H$ and $u=1$. 
The resulting integral can be evaluated to obtain 
\be
\Sigma_0 = {3 \mplan \over 16 \pi R_C^2} 
\left\{ {\sqrt{1 - u} (2 + 3 u) \over u} 
+ 3 u \tanh^{-1} (\sqrt{1 - u}) \right\} \,.
\label{sigmadirect} 
\ee
Once the circumplanetary disk has become well-developed, 
the first term in the above expression dominates, and the 
surface density can be written in the approximate form 
\be
\Sigma_0 \approx {3\mplan \over 8 \pi R_C r} \,.
\label{sigapprox} 
\ee
The enclosed mass thus has the radial dependence $M(r)\propto{r}$. 
As a result, most of the mass enters the planet/disk system at large 
radii $r\gg \rplan$, which in turn implies that disk accretion must 
be crucial for forming planets.\footnote{{\sl Consistency Check:} 
Note that if we integrate the above approximate expression
(\ref{sigapprox}) over the entire extent of the disk, the total mass
is $3\mplan/4$ (which is less than the total mass $\mplan$). We can
show that the full solution provides exact conservation of mass as
follows. The mass in the disk is given by the integral 
$$\mdisk = \int_0^{R_C} 2\pi r dr \Sigma_0 = 
\int_0^{R_C} 2\pi r dr \int_0^T  
{ \dotplan dt \over 4\pi r^2} 
{1 \over (\zeta^2-\zeta)^{1/2}} \,.$$
Let us write the second integral in terms of the 
variable $u = 1/\zeta$ to obtain 
$$\mdisk = {243 M_\ast \over 2 a^3} \int_0^{R_C} r^2 dr 
\int_{u}^1 { du \over u^3 \sqrt{1-u}} = 
{3 \mplan \over 2 R_C^3} \int_0^{R_C} r^2 dr 
\int_{u}^1 { du \over u^3 \sqrt{1-u}} \,. $$
Now switch the order of integration: 
$$\mdisk = {3 \mplan \over 2 R_C^3} 
\int_0^1 { du \over u^3 \sqrt{1-u}} \int_0^{uR_C} r^2 dr = 
{\mplan \over 2} \int_0^1 { du \over \sqrt{1-u}} = \mplan \,. $$
}

Figure \ref{fig:infall} shows how the (projected) infalling
trajectories of the flow join onto the circumplanetary disk. The
inward paths are nearly radial at large distances $r>R_C$ (note that
the length scales in the figure are given in units of the centrifugal
radius $R_C=R_H/3$). In the inner regime, however, the trajectories
depart significantly from radial paths, crossing the midplane and
joining the disk well before reaching the planetary surface. Most 
of the incoming trajectories thus intercept the disk, rather than 
the planetary surface (which corresponds to $x\approx0.006$ in 
the figure). 

For completeness, we note that the solution (\ref{sigmadirect}) 
for the surface density assumes that infall takes place over all
initial polar angles $\theta_0$.  If infall is confined to the polar
regions, as indicated by some numerical treatments (e.g.,
\citealt{lambrechts2017,lambrechts2019}) then the centrifugal barrier
$R_C$ will be smaller and the resulting surface density distribution
will be steeper (see \citealt{fung2019}). This modification,
considering only the range $0\le\theta_0\le\theta_{max}$, is easily
incorporated into this analytic treatment. On the other hand, viscous
evolution will act to spread out the disk, increase its radius, and
make the surface density distribution less steep, as shown in the
following section.

\subsection{Steady State Disk Evolution} 
\label{sec:alphadisk} 

Once the material has fallen onto the circumplanetary disk, its
subsequent evolution occurs through viscous dissipation. Following
standard nomenclature \citep{hartmann2009}, we write the viscosity in
the form
\be
\nu = \alpha c_s H = \alpha c_s^2 / \Omega \,. 
\ee
Note that disk evolution for planet formation lies in a different 
regime than for star formation. The relevant quantity is the 
ratio of the viscous time scale to the accretion time scale, i.e., 
\be
{\cal R} = {R^2 \over \nu} { {\dot M} \over M} = 
{1 \over \alpha} {R^2 \over H^2} { {\dot M} P \over 2\pi M} 
\sim {100 \over 2\pi\alpha} { {\dot M} P \over M} \,,
\label{spratio} 
\ee
where $P$ is the orbit time at the outer disk edge and where the 
final approximate equality assumes $H\sim R/10$. For star
formation, $R\sim100$ AU, $M=M_\ast\sim1M_\odot$, $P\sim1000$ yr, 
and $M_\ast/{\dot M}\sim10^{5}$ yr.  As a result, the ratio 
${\cal R}_{\rm sf}\sim1/2\pi\alpha$. For planet formation,
$\mplan/\dotplan\sim1$ Myr, whereas $P\sim1$ yr, so that 
${\cal R}_{\rm pf}\sim10^{-4}/2\pi\alpha$. A much smaller 
viscosity is required for a planet forming disk to keep up 
with the infall compared to a star forming disk.

The equation of motion for a viscous accretion disk, including the
infall terms derived in the previous section, can be written as 
\be
{\partial \Sigma \over \partial t} = {3 \over r} 
{\partial \over \partial r} \left[ r^{1/2} 
{\partial \over \partial r} \left( r^{1/2} \nu \Sigma \right) 
\right] + {\dotplan \over 4\pi r^2} {1 \over (\zeta^2 - \zeta)^{1/2}} \,.
\label{sigevolve} 
\ee
In steady state, corresponding to vanishing time derivitives, 
the solution takes the from 
\be
\Sigma (r) = { \dotplan \over 6\pi \nu} 
{1 \over \sqrt{u}} \left[ \sin^{-1} \sqrt{u} + 
\sqrt{u (1-u)} \right] = { \dotplan \over 6\pi \nu} f(u) \,,
\label{sssigma} 
\ee
where $u\equiv r/R_C$ and where the second equality defines the
function $f(u)$, which is of order unity and slowly varying with
radius.  Specifically, we note that the function $f\to2$ in the inner
limit $u\to0$ and $f\to\pi/2$ in the outer limit $u\to1$.

The diffusion equation has an additional solution, equivalently, 
an additional term, which has the form 
\be
\nu \Sigma = {K_1 \over r^{1/2}} + K_2 \,,
\ee
where the $K_j$ are integration constants which can be chosen to
satisfy the inner boundary condition. For stellar accretion disks, the
usual assumption is to let $\nu\Sigma\to0$ at the inner boundary. If
we use this inner boundary condition, the full solution can be 
written in the form 
\be
\Sigma (r) = { \dotplan \over 6\pi \nu} \left[ f(u) - f(u_{\rm p}) 
\left({u_{\rm p}\over u}\right)^{1/2} \right] \,,
\label{gensigma} 
\ee
where $u_{\rm p}\equiv \rplan/R_C$. The second term is negligible
except near the surface of the planet. Moreover, in the present
application, the disk is magnetically truncated so that the effective
inner disk edge is well outside the planetary surface (see Section
\ref{sec:magradius}).

We can integrate the steady-state surface density over the 
disk area to find the disk mass, 
\be
\mdisk = {\dotplan\over3}  \int_0^{R_C} {rdr\over\nu} 
{1 \over \sqrt{u}} \left[ \sin^{-1} \sqrt{u} + 
\sqrt{u (1-u)} \right] \,.
\ee
If we assume that the temperature within the disk scales as 
$T\sim r^{-1/2}$, then $\nu\sim r$, so that we can write 
$\nu = \nu_C (r/R_C) = \nu_C u$. The disk integral then becomes
\be
\mdisk = {\dotplan R_C^2\over3\nu_C}  \int_0^1 
{du \over \sqrt{u}} \left[ \sin^{-1} \sqrt{u} + 
\sqrt{u (1-u)} \right] = {\dotplan R_C^2\over3\nu_C}  
\left[ \pi - {4\over3} \right] \,.
\label{diskmass} 
\ee
We thus find that $\mdisk/\mplan\sim {\cal R}$ from equation 
(\ref{spratio}) as expected.

With the steady-state solution in place, we can now consider how the
disk evolves into such a configuration. The steady-state disk arises
when the two terms on the right hand side of equation
(\ref{sigevolve}) are in balance. At the onset of disk formation,
however, the viscous term is small (compared to the infall term)
because the surface density must be small. In this regime, the surface
density of the disk builds up mass from infall as described in the
previous section. The surface density will grow until it becomes
comparable to the steady-state solution, where the required time for
this transient growth is determined by viscosity so that
$\Delta{t}\sim{R_C^2}/\nu_C$. Since the viscous time scale is much
shorter than the evolutionary time scale (equation [\ref{spratio}]),
the disk surface density has the steady-state form for most of its
lifetime. For completeness, note that the centrifugal radius $R_C$
grows with increasing planet mass with the scaling law
$R_C\sim\mplan^{1/3}$. The weak dependence of $R_C$ on mass implies
that $R_C$ varies only weakly with time, so that the evolution of disk
surface density can be separated into the regimes described above.

\begin{figure}
\includegraphics[scale=0.70]{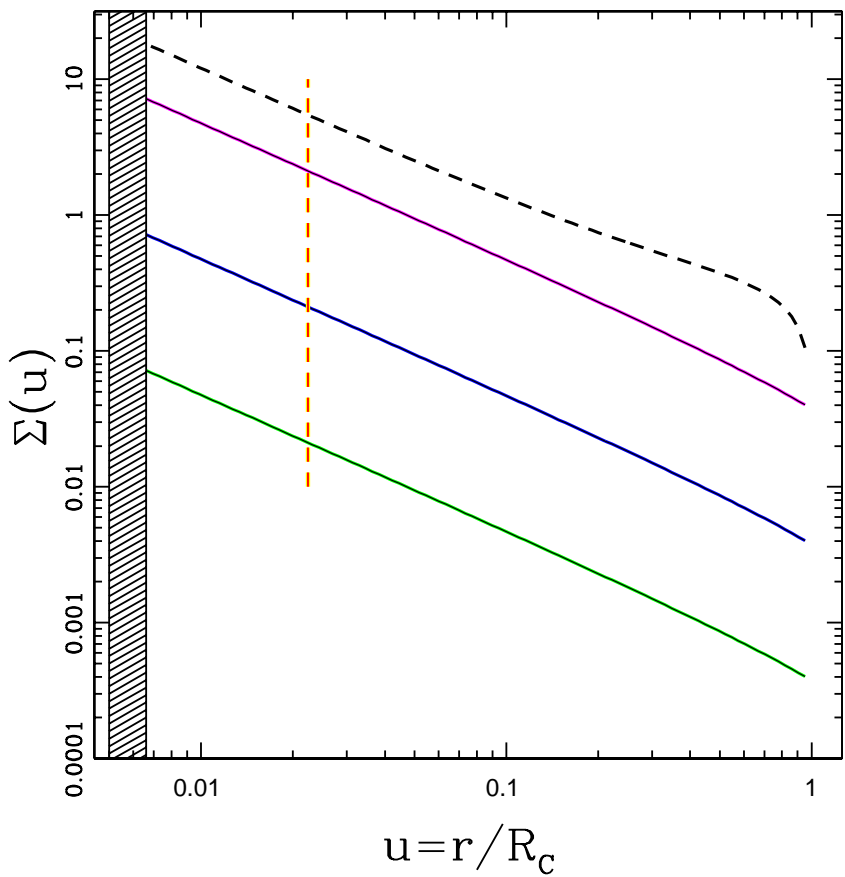}
\vskip-1.50truein
\caption{Surface density distributions for circumplanetary disks 
forming through infall from the circumstellar disk. The black 
dashed curve shows the normalized surface density profile in the 
absence of viscous evolution, where infalling material is assumed 
to stay at the radial location where it crosses the midplane. The 
three solid curves show the normalized steady-state surface 
density distributions for viscous parameters $\alpha$ = $10^{-4}$
(upper, magenta), $10^{-3}$ (middle, blue), and $10^{-2}$ (lower,
green). The hatched region on the left represents the planetary 
surface. The vertical dashed orange line delineates the expected
location of the magnetic truncation radius and hence the effective
inner boundary of the disk. }
\label{fig:sigma} 
\end{figure}

\subsection{Magnetic Truncation at the Inner Boundary} 
\label{sec:magradius} 

With the outer boundary of the disk specified, we need to consider the
inner boundary. In the absence of planetary magnetic fields, the
circumplanetary disks can extend all the way to the planet's
surface. In general, however, we expect the planet to develop a strong
internal magnetic field with surface strength $\bplan$ (see Appendix 
\ref{sec:magfield} for an estimate). To leading order, such fields 
have a dipole form,
\be
B(r,\theta) = \bplan \left({\rplan \over r}\right)^3 
\left[ 3 \cos\theta {\hat r} - {\hat z} \right]\,, 
\label{bdipole} 
\ee
where the spherical coordinate system is centered on the planet. 
The incoming flow from the circumplanetary disk cannot penetrate 
past the point where the ram pressure of the inward flow is 
balanced by the outward pressure from the magnetic field. This 
magnetic truncation radius has the form 
\be
R_X = \omega \left( 
{\bplan^4 \rplan^{12} \over G \mplan \dotplan^2} \right)^{1/7} \,,
\label{rmag} 
\ee
where $\omega$ is a dimensionless constant of order unity
(\citealt{ghosh1978,blandford1982}; see also the discussion 
of \citealt{mohanty2008} for an alternate derivation). 

If we insert typical numbers, the truncation radius can be 
evaluated to find: 
\be
{R_X \over \rplan} \approx 3.8 \left({\mplan\over M_J}\right)^{-1/7} 
\left({\dotplan\over 1\,M_J\,/\,{\rm Myr}}\right)^{-2/7} 
\left({\bplan\over500\,\,{\rm gauss}}\right)^{4/7} 
\left({\rplan\over10^{10}\,\,{\rm cm}}\right)^{5/7} \,. 
\ee
For comparison, the centrifugal radius, and hence the disk radius, 
is given by
\be
\rdisk = R_C = {R_H \over 3} \qquad \Rightarrow \qquad 
{\rdisk\over\rplan} \approx 170 \left({\mplan\over M_J}\right)^{1/3} 
\left({a \over 5 {\rm AU}}\right)\,.
\ee
Since the disk radius scales as with the semimajor axis, planets
forming in the inner regions of the circumstellar disk have radially
small infall regions and produce radially small circumplanetary
disks. For standard parameters, the magnetic truncation radius is
larger than the disk radius for planets forming at semimajor axes
$a\lta0.1$ AU. As a result, magnetic fields could represent a significant
obstacle to forming giant planets near their host stars.\footnote{On the 
other hand, the strong magnetic fields and significant ionization in
this region could lead to efficient magnetic braking and hence another
mechanism to transfer angular momentum.} For completeness, we can write
the condition $R_X=R_C$ in the form 
\be
\left({a\over\rplan}\right) 
\left({\mplan\over M_\ast}\right)^{10/21} = 3^{4/3} \omega
\left( {\bplan^4 \rplan^5 \over GM_\ast \dotplan^2}\right)^{1/7}\,.
\ee

\subsection{Disk Structure} 
\label{sec:disksummary} 

The previous subsections define the expected structure for the
circumplanetary disks that arise during the process of planet
formation. The surface densities for these disks are shown in Figure
\ref{fig:sigma}. The upper black dashed curve shows the surface
density profile that would result in the absence of viscous accretion
(as derived in Section \ref{sec:infalldisk}). The surface density
shows a radial dependence of the approximate form $\Sigma\propto1/r$,
except near the outer edge where an enhancement is present, followed
by a sharp edge at $r=R_C$. In the presence of viscosity, the disk surface
density is given by the steady-state solutions of Section
\ref{sec:alphadisk}. Surface density profiles are shown in the figure
for $\alpha$ = $10^{-2}$, $10^{-3}$, and $10^{-4}$, from bottom to
top. The larger viscosity results in lower disks masses in steady
state. The surface density profiles are normalized by dividing out 
a factor of $\mplan/R_C^2$, where $\mplan$ represents the total mass 
that has fallen to the planet/disk system. For the viscous models, 
we further specify the mass infall rate to be $1M_J$/Myr and the 
radial location of the planet to be $a$ = 5 AU. With these choices, 
for a planet with mass $\mplan=1M_J$, the circumplanetary disk has 
outer radius $R_C/\rplan=170$ and an inner boundary given by the 
magnetic truncation radius at $R_X/\rplan=3.8$, as depicted by 
the vertical orange dashed line in Figure \ref{fig:sigma}. Although
the surface density profiles are truncated at $r=R_C$ in the figure,
the solutions can be extended beyond this nominal disk radius where
they match onto more steeply decreasing profiles (see Appendix
\ref{sec:diskextend}). Note that the disk {\it must} spread beyond 
$R_C$ in order to satisfy conservation of angular momentum.

The circumplanetary disks thus have relatively simple forms.  The
dynamic range is small, with $R_C/R_X\sim40-50$. The disk masses are a
small fraction of the total mass (equation [\ref{diskmass}]) and the
surface density distribution is close to a $1/r$ power-law (equation
[\ref{sssigma}]).  In spite of the small disk masses, the surface
densities are large enough for the disks to remain optically thick to
their internal infared radiation.

\subsection{Planet Masses without Disk Accretion}
\label{sec:noaccretion}  

Before leaving this section, it is useful to consider how gaseous
planets would form in the absence of disk accretion (through the
circumplanetary disk). If the viscosity vanishes, $\nu\to0$, then the
rate at which the planet gains mass is limited by the geometry of the
infall and has the form 
\be
\dotplan\Big|_{direct} = 4\pi \rplan^2 \int_0^1 \rho|v_r|d\mu  \,, 
\ee
where the integrand is evaluated at the planetary surface. Using 
the infall solution from Section \ref{sec:infall} (see also 
\citealt{adams1986} for the star formation analog), we can 
evaluate the integral to obtain 
\be
\dotplan\Big|_{direct} 
= \dotplan \left[ 1 - \sqrt{1-\rplan/R_C} \right] \,.
\label{mdotdirect} 
\ee
The rate of accretion onto the planet is thus an ever-decreasing 
fraction of the total infall rate. 

The planet mass $\mplan$ for a given mass $M$ that has fallen is given
by the integral expression 
\be
\mplan = M_0 + \int_{M_0}^M dM 
\left[ 1 - \sqrt{1-(M_0/M)^{1/3}} \right] \,, 
\label{massint} 
\ee
where the scale $M_0$ is the mass for which the centrifugal 
radius is equal to the planetary radius, i.e., 
\be
M_0 = 81 M_\ast \left({\rplan \over a}\right)^3\,.
\ee
For planets forming at $a\sim5$ AU, $M_0 \sim 2\times10^{-7}$ $M_J$.
In contrast, for hot Jupiters at $a = 0.05$ AU, $M_0\sim0.2M_J$. For
both scenarios, the formation of a Jovian-mass planet requires
accretion via a circumplanetary disk. For cold Jupiters, the vast
majority of the mass must be accreted through the disk. For hot
Jupiters, however, a significant fraction of the mass can be accreted
directly (notice also that hot Jupiters tend to have somewhat smaller 
masses). 

Note that the above treatment does not include the effects of
planetary magnetic fields, which act to increase the effective capture
cross section of the growing planet. For a given estimate of the
effective capture radius $R_\sigma$ (see Section \ref{sec:capture}) 
one can replace $\rplan$ with $R_\sigma$ in the above expressions.

A related question is: How much total mass must flow through the 
Hill sphere in order for a given mass to directly accrete onto the 
planet? We can evaluate equation (\ref{massint}) in the asymptotic 
limit $M\gg M_0$ to obtain 
\be
\mplan \to {3\over4} \left(M_0M^2\right)^{1/3}\,.
\label{pfraction} 
\ee
In order to produce a final planet mass of $\mplan=1M_J$ (with no
circumplanetary disk accretion and at radial location $a=5$ AU), we
would need a total mass $M\approx3400M_J=3.4M_\odot$. In other words,
in order for a Jovian mass to fall directly to the planet, several
stellar masses of material must cycle through the Hill sphere. Some
mechanism for redistributing angular momentum --- within the sphere of
influence of the forming object --- is thus necessary for the
formation of Jovian planets. Viscosity in the circumplanetary disk
provides such a mechanism for the scenario explored in this
paper.\footnote{Note that this derivation assumes that the total mass
  accretion rate $\dotplan$ has a spherically symmetric distribution
  for the material that passes through the Hill radius. If other
  geometries are assumed, the exponents and coefficients in equation
  (\ref{pfraction}) change accordingly. The general result --- that
  the total mass flowing into the Hill sphere must be much larger than
  the final mass of the planet --- remains robust. }


\subsection{Magnetic Capture Radius} 
\label{sec:capture} 

The magnetic truncation radius $R_X$ (equation [\ref{rmag}]), which
arises from the balance between disk accretion and magnetic pressure,
defines the inner edge of the circumplanetary disk. However, for
material falling onto the system near the rotational poles, a second
magnetic boundary $R_\sigma$ arises from the balance between the ram
pressure of infalling material and the magnetic pressure. These two
scales are different because not all of the infalling material can
reach inner radii comparable to the planetary size, so that the mass
infall rate is reduced and the boundary is larger. Equating the
incoming ram pressure $\rho v^2$ with the magnetic field pressure at
the pole ($\mu=1$), we can write the magnetic capture radius in the form
\be 
{R_\sigma \over \rplan} = {\tilde\omega} \left( 
{\bplan^4 \rplan^3 R_C^2 \over G \mplan \dotplan^2} \right)^{1/9} \,, 
\label{rcap} 
\ee 
where ${\tilde\omega}$ is a dimensionless constant of order unity and
where the quantity $\dotplan$ is the total mass accretion rate.  For
typical parameter values, the capture radius is somewhat larger than
the disk truncation radius so that $R_\sigma\sim1.4R_X$ (compare with
equation [\ref{rmag}]). 

\begin{figure}
\vskip-0.3truein
\includegraphics[scale=0.70]{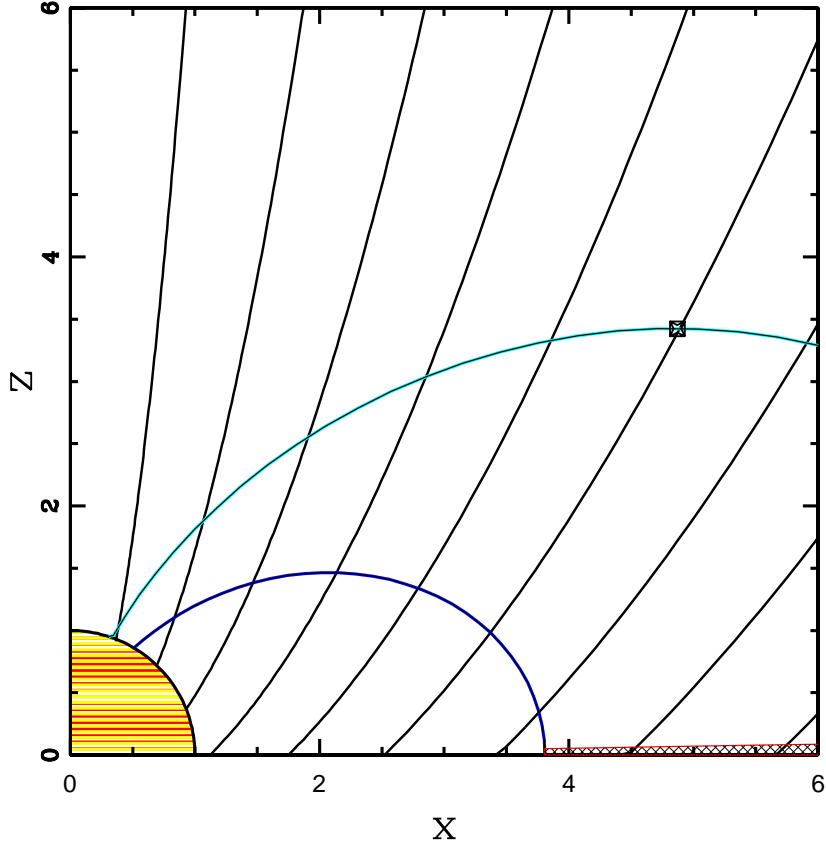}
\vskip-1.60truein
\caption{Inner infall region near the planet (for mass $\mplan=1M_J$, 
accretion rate $\dotplan=1M_J$/Myr, and field strength $\bplan=500$
gauss).  The black solid curves show the projected incoming
trajectories for parcels of gas with starting conditions near the
pole. The planet is depicted as the (partial) circular region in the
lower left. Two magnetic field lines are shown, with the lower blue
curve connecting to the magnetic truncation radius $R_X$ of the disk
(in the equatorial plane) and the upper cyan curve defining the
capture radius $R_\sigma$, (marked by the square symbol). The
horizontal point of the upper curve, $x_{\rm hor}=R_\sigma\sin\theta$,
is roughly coincident with the disk truncation radius defined by the
lower curve.  The length scales are given in units of the planetary
radius $\rplan$, so that the disk radius $\rdisk=R_C\approx180$ and
the Hill radius $R_H\approx540$. }
\label{fig:capture} 
\end{figure}

The magnetic field lines and incoming trajectories near the forming
planet are shown in Figure \ref{fig:capture}. Keep in mind that all of
the trajectories in the figure correspond to small values of the
initial polar angle $\theta$ (otherwise the trajectories would cross
the plane of the disk far to the right). In order for an incoming
parcel of gas to be deflected by the magnetic fields, its radial
location is given by equation (\ref{rcap}). In order for the parcel to
subsequently move inward onto the planet, rather than outward and back
towards the disk, it must fall sufficiently close to the pole. Here we
approximate this bifurcation point as the point where the magnetic
field line is horizontal and where the magnetic pressure balances the
ram pressure. This critical point of the field line corresponds 
to the location where the horizontal coordinate has the form 
\be
x_{\rm hor} = \left({2\over3}\right)^{1/2}R_\sigma \sim 0.82 R_\sigma \,.
\ee
We thus have the following result: The capture radius $R_\sigma$ is
larger than the magnetic truncation radius because the direct infall
is reduced in the inner region. However, only the innermost 
portion of the capture region funnels material directly onto the
planet rather than outward and onto the disk (see
\citealt{batygin2018,batmorby2020}).  The radius $R_\sigma$ is
somewhat larger than $R_X$ and $x_{\rm hor}$ is somewhat smaller than
$R_\sigma$, so that the two corrections tend to compensate. In the 
end, the effective capture cross section is given approximately by the
length scale $R_X$.

\section{Radiative Signatures} 
\label{sec:radiation} 

With the properties of the infalling envelope and the circumplanetary 
disk specified, we can now estimate the accretion luminosity of the 
central planet/disk system and the corresponding spectral energy 
distributions. 

\subsection{Luminosity Contributions from the Planet and its Disk}
\label{sec:luminosity} 

Since most of the mass falling onto the planet/disk system is
eventually accreted onto the growing planet, the luminosity
contributions can be written in terms of the benchmark power scale
\be
L_0 \equiv {G \mplan \dotplan \over \rplan} \,, 
\label{lumzero} 
\ee
which corresponds to the total power that could be generated if all of
the incoming material falls (from rest at infinity) onto the planetary
surface at free-fall speeds and dissipates all of its kinetic energy.

The luminosity of the system is distributed between the planet and
its disk. Only a fraction $f_{\rm p}$ of the incoming material falls
directly onto the planet (see equation [\ref{mdotdirect}]). However,
an additional fraction of the material falls within the magnetic
capture radius (Section \ref{sec:capture}) and is funneled onto the
planet along magnetic field lines. The remaining incoming
material falls onto the disk. For the sake of definiteness and
simplicity, we assume that the magnetic truncation radius $R_X$ of the
disk and the magnetic capture radius $R_\sigma$ are equal. The
fractions of the incoming material that fall onto the planet and onto
the disk are thus given by 
\be
f_{\rm p} = 1 - \sqrt{1 - \ups} \qquad {\rm and} \qquad 
f_{\rm d} = \sqrt{1 - \ups} \,,
\ee
where we have defined $\ups\equiv R_X/R_C$. If the planetary 
magnetic field is sufficiently weak, so that $R_X\le\rplan$,
then $\ups=\rplan/R_C$.

To specify the planetary component of the accretion luminosity, we
assume that all of the material that falls within the capture radius
is accreted by the planet, and that shocks on the surface dissipate
the radial and poloidal components of the velocity.  The planet is
assumed to be co-rotating with its disk at the magnetic truncation
point, so that the planetary rotation rate is given by 
\be
\Omega_{\rm p}^2 = {G \mplan \over R_X^3} \,. 
\label{omegaplan} 
\ee
The incoming material thus retains a small azimuthal velocity given 
by $v_\phi = \sin\theta\rplan\Omega_{\rm p}$. The luminosity thus 
takes the form 
\be
L_{\rm p}^{\rm dir} = \left(1 - \sqrt{1 - \ups} \right) 
\left( 1 - {1\over2}\langle\sin^2\theta\rangle {\rplan^3\over R_X^3} 
\right) L_0 \,, 
\label{lplansurf} 
\ee
where the angular brackets represent an average over the co-latitudes. 
The newly added material is eventually distributed over the planetary 
surface, so we expect $\langle\sin^2\theta\rangle\approx2/3$. 

Although the remaining fraction of the material falls onto the disk,
it eventually accretes onto the planet. However, it falls from the
magnetic truncation radius, rather than from infinity, so that the
additional planetary luminosity derived from accretion takes the
approximate form 
\be
L_{\rm p}^{\rm acc} = L_0 \sqrt{1-\ups} 
\left( 1 - {\rplan\over R_X} \right) 
\left( 1 - {\rplan^3\over 3R_X^3} \right) \,.
\label{lplanrx} 
\ee
As written, the expression has three correction factors: The first
determines the amount of material that falls onto the disk (the
remaining fraction is accounted for in the previous contribution). The
second factor takes into account the starting point for the material,
which falls from an initial radius $R_X$ instead of $r\to\infty$. The
third factor takes into account the rotational motion of the incoming
material, where we again assume that the magnetic field causes the
incoming material to rotate with the mean motion of the disk at the
location $R_X$ and that the incoming material is uniformly distributed
in planetary latitude. Although this correction factor is close to
unity, note that other assumptions can be invoked. For completeness, 
we note that the planet and its disk could support outflows or winds
analogous to those observed in young stellar objects. Such winds could
result in a reduced accretion rate onto the planetary surface, and 
could play a role in redistributing angular momentum. Although 
potentially important, we leave this issue for future work. 

The luminosity contributions from both direct infall and from the
inner edge of the disk combine to determine the total planetary
accretion luminosity $L_{\rm p}$, which takes the form 
\be
L_{\rm p} = \left( 1 - \sqrt{1-\ups} {\rplan\over R_X} \right) 
\left( 1 - {\rplan^3\over 3R_X^3} \right) L_0 \,.
\label{lplantot} 
\ee
In general, the planet has an additional internal luminosity 
$L_{\rm int}$, although the accretion luminosity is expected to be
somewhat larger (see, e.g., \citealt{marley2007}). The planet must
radiate the total luminosity, given by equation (\ref{lplantot}) and
$L_{\rm int}$, and has an effective surface temperature given by
\be
\sigma T_{\rm p}^4 = {L_0 \over 4\pi \rplan^2} 
\left( 1 - \sqrt{1-\ups} {\rplan\over R_X} \right) 
\left( 1 - {\rplan^3\over 3R_X^3} \right) + 
{L_{\rm int} \over 4\pi\rplan^2} \,. 
\label{plantemp} 
\ee
In the limit $L_{\rm int}\to0$, only the first term contributes, 
and we make this approximation for the rest of this paper. 

The remainder of the energy is dissipated within the disk, 
which has total luminosity given by 
\be
L_{\rm d} = L_0 {\rplan\over 2R_X} \sqrt{1-\ups} \,,
\label{ldisktotal} 
\ee
where the factor of 2 arises because half of the energy is stored in
rotational energy of the disk. Because the magnetic truncation radius
is typically $R_X\sim4\rplan$, the total accretion luminosity
generated on the planetary surface is larger than the disk luminosity
by (about) an order of magnitude. From the perspective of the
infalling envelope, most of the energy is generated by a point source
in the center of the structure.

The total disk luminosity is distributed among three components: 
The perpendicular component $v_\theta$ of the incoming velocity is
dissipated as it strikes the disk, thereby producing luminosity
$L_{\rm d}^{\rm surf}$.  When the material reaches the disk, however, it
generally will not have the right azimuthal velocity to enter into
Keplerian rotation about the planet. The new material thus mixes with
existing disk material, dissipates energy while conserving angular
momentum, and produces a mixing luminosity $L_{\rm d}^{\rm mix}$. The
remaining luminosity $L_{\rm d}^{\rm visc}$ must be dissipated (in this
approximation) by the viscosity. The surface and mixing contributions
for circumstellar disks have been calculated previously 
\citep{cassen1981,cassen1983,adams1986} and can be written in the form 
\be
L_{\rm d}^{\rm surf} = \int_A \left({1\over2} v_\theta^2 \right) 
\rho |v_\theta| dA = {\ups \over 4} \left\{ \ln \left[
{1 + \sqrt{1-\ups} \over 1 - \sqrt{1-\ups}}\right] - 2\sqrt{1-\ups} 
\right\} L_0 \,,
\label{ldisksurf} 
\ee
and
$$
L_{\rm d}^{\rm mix} = \int_A {1\over2} \left\{ v_r^2 + 
\left[v_\phi - \left({G\mplan\over r}\right)^{1/2}\right]^2 \right\} 
\rho |v_\theta| dA 
$$
\be
= {\ups \over 2} \left\{ \ln \left[
{1 + \sqrt{1-\ups} \over 1 - \sqrt{1-\ups}}\right] + \sqrt{1-\ups} 
+ \pi - 2 \tan^{-1} \left[ {\ups \over 1-\ups}\right]^{1/2} 
\right\} L_0 \,. 
\label{ldiskmix} 
\ee
These two luminosity contributions (see also \citealt{szulagyimord})
represent the energy that is dissipated as material joins the disk,
independent of any subsequent viscous evolution, and can be combined
to define a total direct luminosity contribution $L_{\rm d}^{\rm dir}$
= $L_{\rm d}^{\rm surf}+$ $L_{\rm d}^{\rm mix}$, i.e., 
\be
L_{\rm d}^{\rm dir} = {\ups \over 4} \left\{ 3 \ln \left[
{1 + \sqrt{1-\ups} \over 1 - \sqrt{1-\ups}}\right] + 2\pi - 4 
\sin^{-1} \sqrt{\ups} \right\} L_0 \,.
\label{ldiskdir}  
\ee 
In the limit where all of the material can be accreted onto the 
planet, the remaining viscous luminosity is given by  
\be
L_{\rm d}^{\rm visc} = L_{\rm d} - L_{\rm d}^{\rm dir} \,.
\label{ldiskvis} 
\ee
Although we can separate the contributions to the disk luminosity, the
net result of viscous evolution is to transport all of the incoming
material inward to the magnetic truncation radius. As a result, the
total disk luminosity has the simple form given by equation
(\ref{ldisktotal}). We can approximate the temperature distribution 
from the disk surface using the form 
\be
T_{\rm d} (r) = T_X \left({R_X \over r}\right)^{3/4}\,,
\label{tempdisk} 
\ee
where the leading coefficient is determined by the requirement 
that the disk radiates its total luminosity, 
\be
\sigma T_X^4 = {G \mplan \dotplan \over 8\pi R_X^3}
{1 \over \sqrt{1 - \ups}} \,.
\ee
Note that the power-law dependence of the temperature distribution 
(\ref{tempdisk}) does not continue to arbitrarily large radii. The 
forming planet is embedded within the circumstellar disk of its 
host star, and this nebula provides a background minimum temperature 
for the circumplanetary disk. 

With the surface temperatures of the planet and the disk specified, 
we can define the corresponding spectral energy distributions of 
the two central components
\be
L_{{\rm p}\nu} = 4\pi \rplan^2 \pi B_\nu(\tplan) \,,
\ee
and 
\be
L_{{\rm d}\nu} = 2 (2\cos\theta) 
\int_{R_X}^{R_C} \pi B_\nu[T_{\rm d}(r)] 2\pi rdr \,.
\ee
The leading factor of 2 arises because the disk has two sides and the
factor of $(2\cos\theta)$ arises from the viewing angle. Here we have
assumed that the disk is optically thick to its own emission and this
expression does not include external extinction (see also
\citealt{zhu2015,zhu2018}).  The corresponding monochromatic fluxes 
arising from the planet and its disk have the form  
\be
F_{{\rm p}\nu} (r,\mu) = {L_{{\rm p}\nu} \over 4\pi r^2} 
G_{\rm pd} (\mu) \exp[-\tau_\nu (r,\mu)] \,, 
\ee
and 
\be
F_{{\rm d}\nu} (r,\mu) = {L_{{\rm d}\nu} \over 4\pi r^2} 
(2\mu) G_{\rm dp}(\mu) \exp[-\tau_\nu (r,\mu)] \,, 
\ee
as seen by an observer with location $(r,\mu)$. The optical depth of
the infalling envelope $\tau_\nu$ depends on the viewing angle. 
For completeness, we include the functions $G_{\rm dp}(\mu)$ and
$G_{\rm pd}(\mu)$, which take into account the mutual shadowing of the
planet by the disk, and the disk by the planet \citep{adams1986}. Most
of the shadowing takes place near the planet, however, and we expect
the magnetic truncation radius to remove the nearest disk material, so
that the shadowing factors are close to unity in the present
application.

\subsection{Radiation from the Circumplanetary Envelope} 
\label{sec:envelope} 

Here we consider the infalling envelope surrounding the forming planet
to be optically thin to its own emitted thermal photons. In this limit,
the spectral energy ${\cal L}_\nu$ corresponding to dust emission has 
the form
\be
{\cal L}_\nu = f_e \int_{R_X}^{R_H} \rho \kappa_\nu  
4\pi B_\nu(T)\,4\pi r^2 dr \,, 
\ee
where $\kappa_\nu$ is the frequency dependent opacity and $B_\nu$ is
the Planck function. The magnetic truncation radius $R_X$ delineates
the inner boundary of the envelope and the Hill radius $R_H$ marks the
outer boundary of the infall region.  We are thus implicitly assuming
that the radiation field reaches its eventual free-streaming form for
$r \lta R_H$. The factor $f_e\le1$ represents a covering fraction,
which allows for a possible reduction in the solid angle subtended by 
the envelope, with a corresponding reduction in the emitted radiation
(although we take $f_e=1$ for the sake of definiteness). The density
distribution is that found in Section \ref{sec:infall}, where we use 
the equivalent spherical distribution for simplicity.\footnote{Note 
that in general the density distribution will depend depend on both
angles $(\theta,\phi)$. This complication is beyond the scope of 
this present treatment, but should be considered in future work.} 

The interaction between the radiation field and the envelope is
dominated by the dust opacity $\kappa_\nu$. Although the dust opacity
in the interstellar medium has been well characterized (starting with
\citealt{draine1984}), the dust within circumstellar disks is expected
to evolve significantly from its interstellar form. Dust grains
generally grow, and opacity decreases as the grain size becomes
comparable to the wavelength of the radiation. In the present context,
the temperature of the envelope falls in the range 100 K $\le{T}\le$
1000 K. Over this interval, the Planck mean and Rosseland mean
opacities are slowly varying functions of temperature, and the
different possible models are in relatively good agreement (e.g., see
\citealt{semenov2003} and references therein). As a result, to a good
approximation, we can consider the opacity to have the power-law form
$\kappa_\nu=\kappa_0 (\nu/\nu_0)^\kapindex$. The Planck mean opacity
then has the temperature dependence $\kappa_P=\bkap T^{\,\kapindex}$,
where the constant $\bkap$ is given by
\be 
\bkap \equiv \kappa_0 \left({k \over h\nu_0}\right)^\kapindex
{\Gamma(4+\kapindex)\,\zetarm(4+\kapindex))\over6\,\zetarm(4)} \,.  
\ee 
Here $k$ and $h$ are the Boltzmann and Planck constants, respectively,
$\Gamma(z)$ is the gamma function and $\zetarm(z)$ is the Riemann zeta
function \citep{abrasteg}.  With the choices $\kapindex=1$, $\kappa_0$
= 10 cm$^2$/g, and $\nu_0=10^{14}$ Hertz, the resulting $\kappa_\nu$
is somewhat smaller than the interstellar opacity, but the resulting
mean opacities are within the ranges advocated by \cite{semenov2003}
for circumstellar disks.  We use these values for the results
presented below, but they are readily varied given the analytical
solutions developed herein.

As a further simplification, we note that the infalling envelope is
optically thin to its emitted radiation field. In this regime, we can
solve the radiative transfer equation \citep{adams1985} and find that 
the temperature profile takes the power-law form 
\be 
T(r) = T_C \left({r\over R_C}\right)^{-2/5} \,, 
\ee
where the power-law exponent is appropriate for the dust opacity law
($\kappa_\nu\propto\nu$) used above (in general, the exponent is
$2/(\kapindex+4)$). With the temperature profile specified, we can 
evaluate the total luminosity emitted by the infalling envelope to find
\be 
{\cal L} = 16\pi \int_{R_X}^{R_H} \rho \kappa_P
\sigma T^4 r^2 dr \,, 
\ee 
where $\kappa_P$ is the Planck mean opacity. The integral expression 
for the envelope luminosity ${\cal L}$ can be evaluated and has the form
\be 
{\cal L} = 16\pi \,\sigma\, C \bkap T_C^5 R_C^{3/2} 
\int_{\ups}^{3} u^{-3/2} {\cal A}(u) du = 
16\pi R_C^2 \,\sigma\, \bkap T_C^5 N_{\rm col} \,.  
\label{envelint} 
\ee 
Note that the upper limit of the integral in equation (\ref{envelint})
is taken to be the outer boundary of the infalling envelope at $u=3$
(equivalently, $r=R_H$).\footnote{In principle, one could extend the 
integral out to $u\to\infty$. In that case, the temperature $T_C$
would be that required to produce the luminosity ${\cal L}$ over the
entire volume, and one would have to evaluate the spectral energy
distribution at a large radius, instead of the outer boundary at
$r=R_H$. Moreover, for $r>R_H$, the geometry and other properties 
of the background circumstellar disk will post-process the spectral 
energy distribution, so that the result will depend on the 
environment. } The fiducial temperature $T_C$ is thus given by
\be 
T_C = \left( {{\cal L} \over 
16\pi R_C^2 \sigma N_{\rm col} \bkap}\right)^{1/5} \,.  
\label{tdust} 
\ee 
The total envelope luminosity ${\cal L}$ is determined by the 
fraction of the central source radiation that is attenuated by 
the infalling envelope, 
\be 
{\cal L} = \left[ L_{\rm p} + L_{\rm d} - 
\int_0^\infty d\nu (L_{{\rm p}\nu} + 
L_{{\rm d}\nu}) \exp(-\tau_\nu) \right] \,, 
\ee 
where $\tau_\nu$ = $\kappa_\nu N_{\rm col}$ is the total optical depth
through the envelope for a given frequency.  For typical parameters,
the fraction of the central source luminosity that is absorbed by the
envelope is $\sim0.25-0.5$, i.e., ${\cal L}\sim0.25$ $(L_{\rm p} +
L_{\rm d})$. The optical depth of the envelope to the central source 
radiation is thus of order unity. The optical depth of the envelope 
to its emitted radiation, which has longer wavelength, is much less
than unity, so that the optically thin approximation is valid.

\begin{figure}
\includegraphics[scale=0.70]{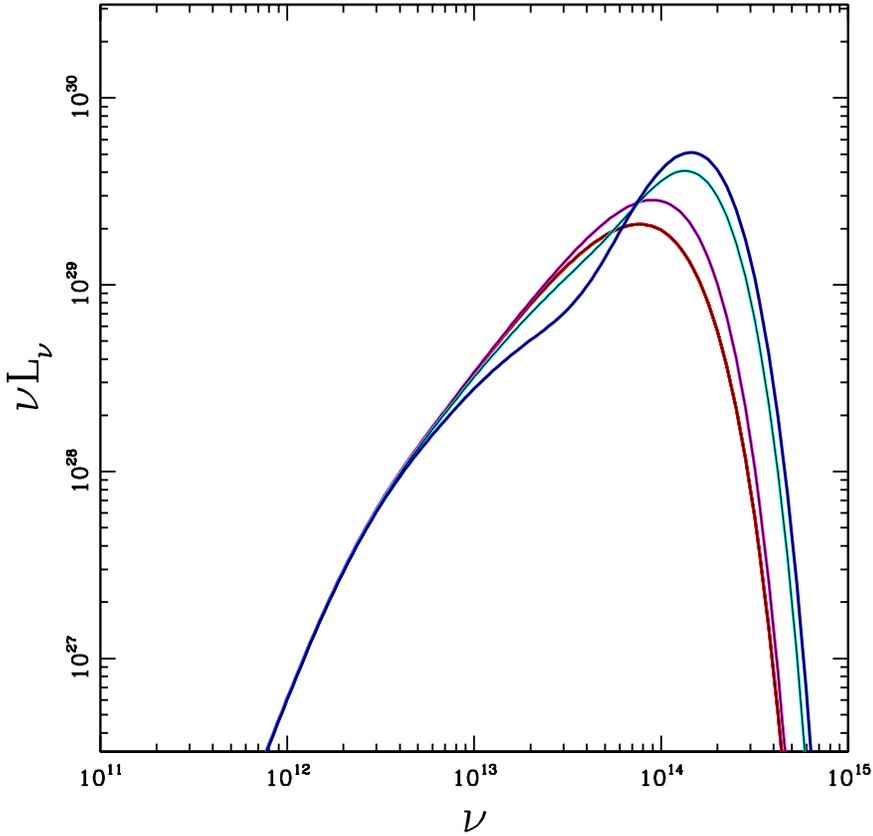}
\vskip-1.50truein
\caption{Spectral energy distributions of the central planet and its 
circumplanetary disk. The planet has mass $\mplan=1M_J$, mass
accretion rate $\dotplan$ = $1M_J$/Myr, and radial location $a=5$ AU. 
Results are shown in the absence of extinction for a range of magnetic
truncation radii, from $R_X=R_P$ out to $R_X\approx5.6R_P$,
corresponding to surface magnetic field strengths $B_{\rm p}$ = 10,
$10^{3/2}$, 100, $10^{5/2}$, and 1000 gauss (from bottom to top). The
frequency $\nu$ is given in Hertz, and spectral energy $\nu L_\nu$ is
given in erg/s. }
\label{fig:source} 
\end{figure} 

\begin{figure}
\includegraphics[scale=0.70]{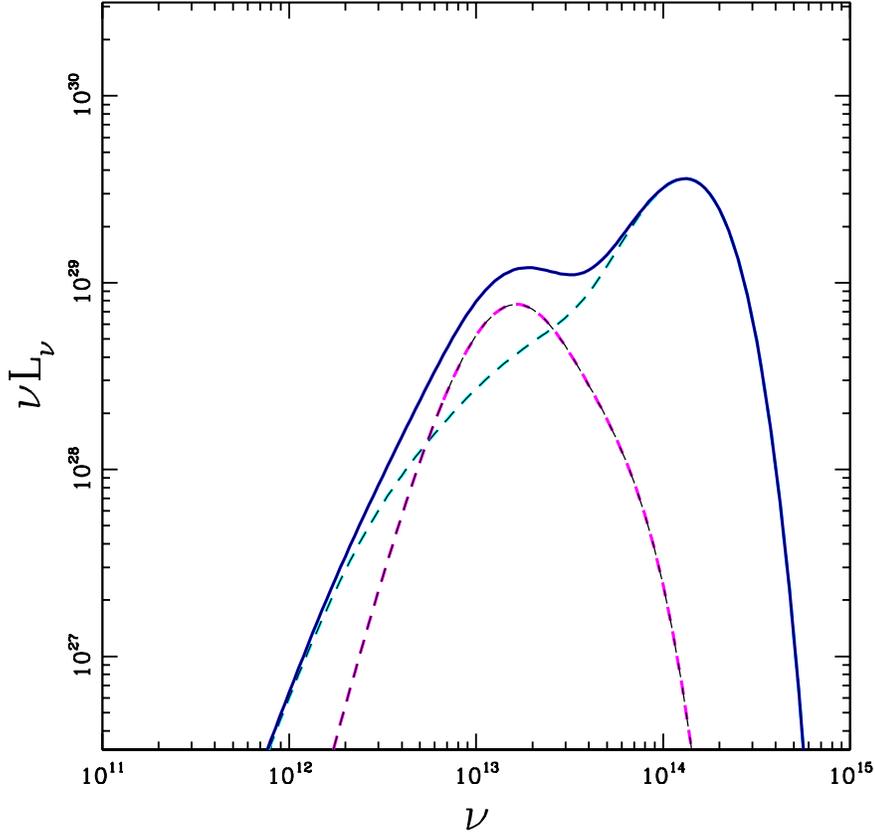}
\vskip-1.50truein
\caption{Emergent spectral energy distributions for a forming planet. 
The planet is assumed to have mass $\mplan = 1 M_J$, radial location 
$a$ = 5 AU, mass accretion rate $\dotplan$ = $1M_J$/Myr, and have 
surface magnetic field strength $B_{\rm p}=1000$ gauss. Emission 
from the central planet and disk is attenuated by the infalling 
envelope, with emergent spectral energy shown as the dashed cyan 
curve. Emission from the envelope itself is shown as the dashed 
magenta curve. The total spectral energy distribution is shown 
as the solid blue curve. (Frequency $\nu$ is in Hertz and spectral 
energy $\nu L_\nu$ is in erg/s.) } 
\label{fig:emergent} 
\end{figure} 

\subsection{Spectral Energy Distributions} 
\label{sec:sed} 

With the properties of the planet, disk, and infalling envelope 
specified, we can determine the spectral energy distribution of 
the forming planet. Here we present only a simple continuum model
where the planetary surface and the disk surface are assumed to 
radiate as blackbodies. Future work should consider line emission. 

The spectral energy distributions of the central planet/disk system
are shown in Figure \ref{fig:source}. The planet mass is taken to be
$\mplan = 1 M_J$ with a mass accretion rate $\dotplan$ = $1M_J$/Myr.
The planet is assumed to form at radial location $a$ = 5 AU, which
determines its Hill radius and hence the disk radius. For a given
choice of magnetic field strength on the planetary surface, the inner
edge of the disk is determined by the magnetic trunaction radius. The
curves shown in the Figure correspond to logarithmically spaced values
in the range $B_{\rm p}$ = 10 -- 1000 gauss (increasing from bottom to
top). Stronger magnetic fields result in larger magnetic truncation
radii $R_X$. Under the assumption of steady state disk accretion,
larger values of $R_X$ result in larger overall luminosity and more
accretion power generated on the planetary surface (see Section 
\ref{sec:luminosity}). Smaller inner radii $R_X$ allow for relatively 
more energy to be dissipated in the disk, rather than on the planetary 
surface, so that the spectral energy distributions become somewhat 
redder. The differences are modest, however, with the total luminosity 
varying by only a factor of $\sim2$ over the given parameter range. 

Radiation from the central planet/disk system is absorbed and
re-radiated by the infalling envelope (Section \ref{sec:envelope}).
The total optical depth of the envelope to the central source photons
is less than unity, so that only a fraction of the total luminosity is
reprocessed. Figure \ref{fig:emergent} shows the resulting emergent
spectral energy distribution for a typical system with mass $\mplan$ =
1 $M_J$, mass accretion rate ${\dot M}$ = 1 $M_J$/Myr, location $a=5$
AU, and magnetic field strength $\bplan=1000$ gauss (so that
$R_X\approx5.6$ $\rplan$). The contributions from the central source
and the envelope are shown separately (dashed curves), with the total
given by the solid curve. As expected, the envelope dominates the
emission at sufficiently long wavelengths ($\lambda\sim10$ $\mu$m),
whereas more energy is emitted at shorter wavelengths ($\lambda\lta1$
$\mu$m).

\begin{figure}
\includegraphics[scale=0.70]{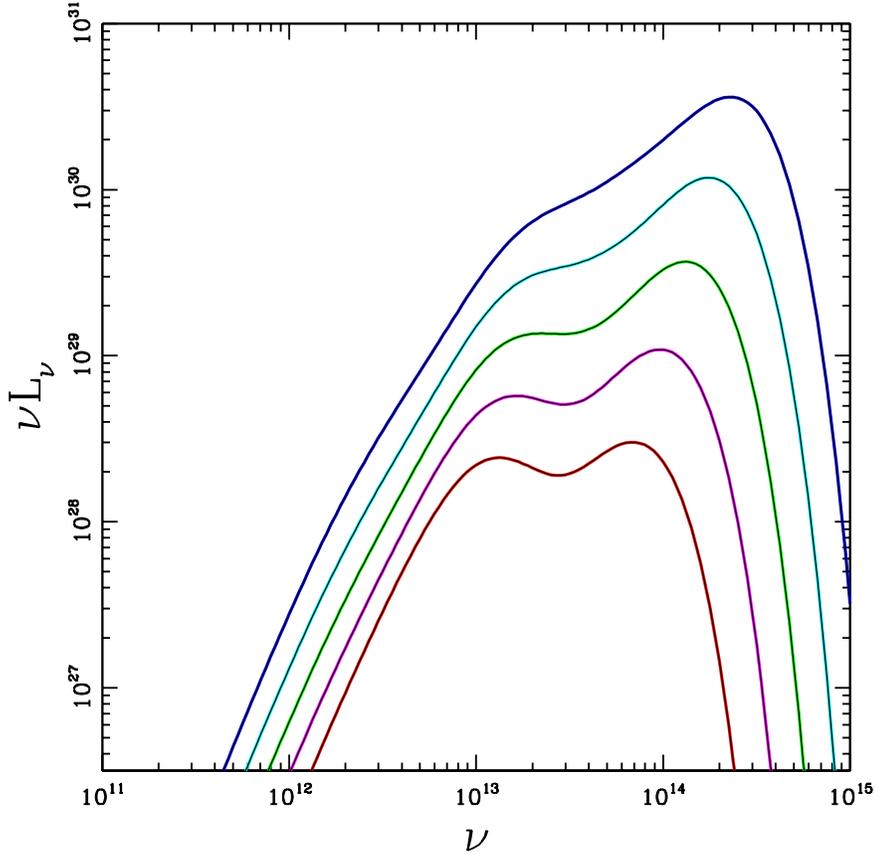}
\vskip-1.50truein
\caption{Emergent spectral energy distributions for a forming planet 
as a function of mass $\mplan$. The planet is assumed to have mass
accretion rate $\dotplan$ = $1M_J$/Myr, radial location $a$ = 5 AU,
and surface magnetic field strength $B_{\rm p}=500$ gauss. The curves
show the spectral energy distributions for planet masses $\mplan$ =
0.1 $M_J$ (bottom, red), $10^{-1/2}M_J$ (magenta), $1M_J$ (middle,
green), $10^{1/2}M_J$ (cyan), and $10M_J$ (top, blue). (Frequency
$\nu$ is in Hertz and spectral energy $\nu L_\nu$ is in erg/s.) }
\label{fig:sedmass} 
\end{figure} 

\begin{figure}
\includegraphics[scale=0.70]{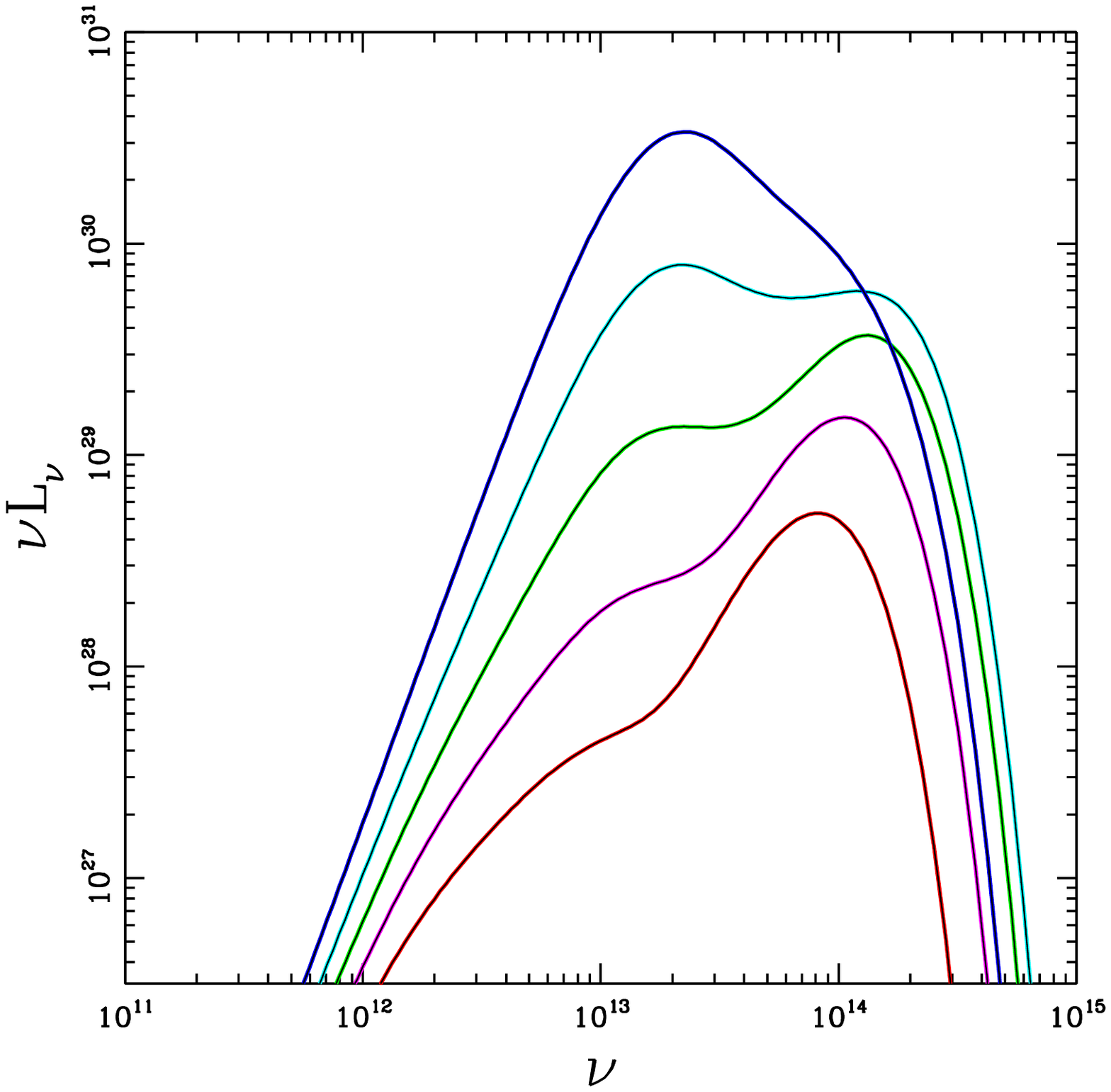}
\vskip-1.50truein
\caption{Emergent spectral energy distributions for a forming planet 
as a function of mass accretion rate $\dotplan$.  The planet is
assumed to have mass $\mplan=1M_J$, radial location $a$ = 5 AU, and
surface magnetic field strength $B_{\rm p}=500$ gauss.  The curves
show the spectral energy distributions for mass accretion rates
$\dotplan$ = 0.1 $M_J$/Myr (bottom, red), $10^{-1/2}M_J$/Myr
(magenta), $1M_J$/Myr (middle, green), $10^{1/2}M_J$/Myr (cyan), and
$10M_J$/Myr (top, blue). (Frequency $\nu$ is in Hertz and spectral
energy $\nu L_\nu$ is in erg/s.) }
\label{fig:sedmdot} 
\end{figure} 

\begin{figure}
\includegraphics[scale=0.70]{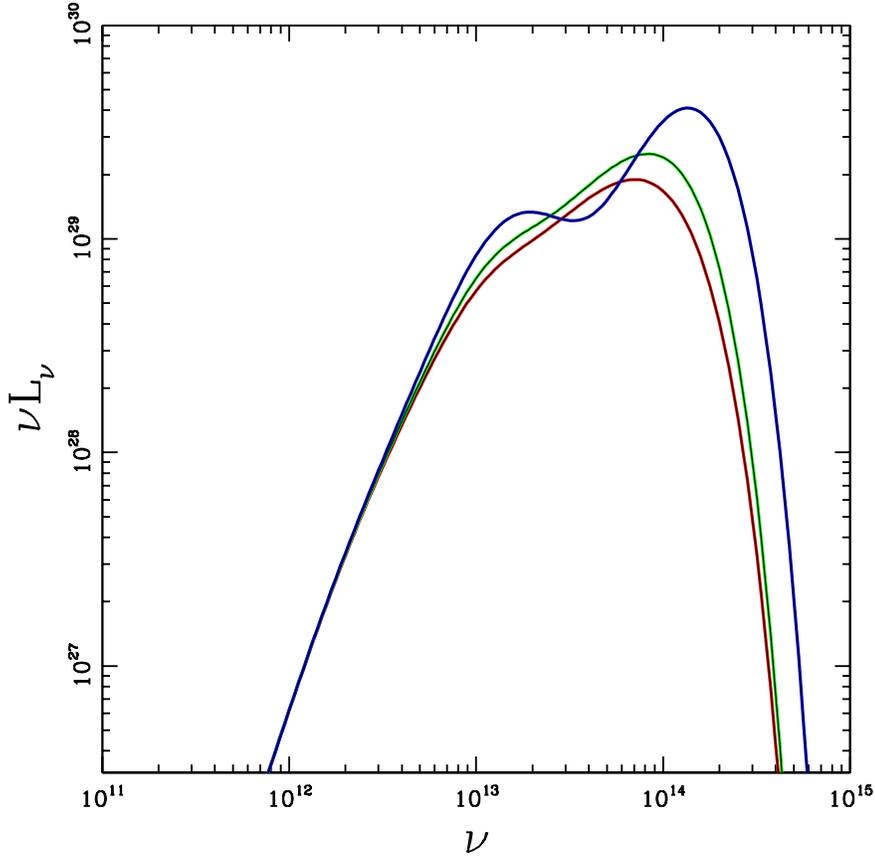}
\vskip-1.50truein
\caption{Emergent spectral energy distributions for a forming planet 
as a function of surface magnetic field strength $\bplan$ (which sets
the location of the magnetic truncation radius $R_X$).  The planet is
assumed to have mass $\mplan=1M_J$, mass accretion rate
$\dotplan=1M_J$/Myr, and radial location $a$ = 5 AU. The curves show
the spectral energy distributions for surface magnetic field strength
$B_{\rm p}=10$ gauss (bottom, red), 100 gauss (middle, green), and
1000 gauss (top, blue). For $\bplan$ = 10 gauss, the disk extends 
to the planetary surface. For $\bplan$ = 100 (1000) gauss, the disk 
is truncated at $R_X\approx1.5(5.6)\rplan$. (Frequency $\nu$ is in 
Hertz and spectral energy $\nu L_\nu$ is in erg/s.) } 
\label{fig:sedbmag} 
\end{figure} 

\begin{figure}
\includegraphics[scale=0.70]{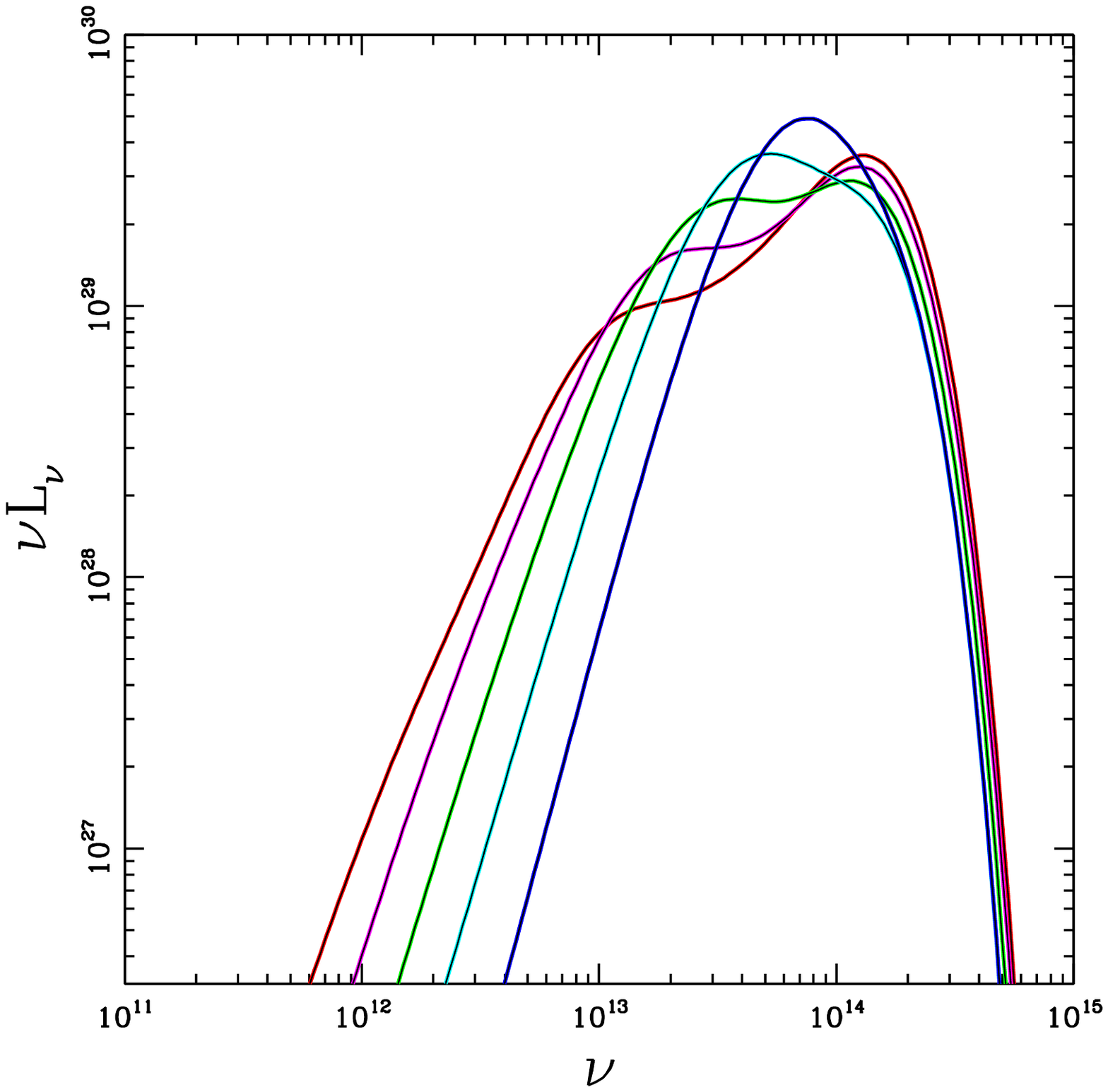}
\vskip-1.50truein
\caption{Emergent spectral energy distributions for a forming planet 
as a function of semimajor axis $a$ of the planetary orbit. The planet 
is assumed to have mass $\mplan=1M_J$, mass accretion rate 
$\dotplan=1M_J$/Myr, and surface magnetic field strength $\bplan$ =
350 gauss.  The curves show the spectral energy distributions for
semimajor axes $a$ = 10 AU (red), $10^{1/2}$ AU (magenta), 1 AU
(green), $10^{-1/2}$ AU (cyan), and 0.1 AU (blue). (Frequency 
$\nu$ is in Hertz and spectral energy $\nu L_\nu$ is in erg/s.) }
\label{fig:sedarad} 
\end{figure} 

The emergent spectral energy distributions depend on four variables
that define the system: the planet mass $\mplan$, the mass accretion
rate $\dotplan$, the magnetic field strength $\bplan$ on the planetary
surface, and the semimajor axis $a$ of the planetary orbit. With
$(\mplan,\dotplan)$ specified, the field strength determines the
magnetic truncation radius, which sets the inner edge of the disk.
The semimajor axis determines the centrifugal radius, which sets the
outer edge of the disk. We note that if a complete theoretical
description of the planet formation process were available, then the
dependence of the mass accretion rate and the magnetic field strength
on planet mass and location could be calculated (see also Appendix 
\ref{sec:evolution}).\footnote{In general, we expect that the 
{\it distributions} of these quantities would be calcuable -- the
evolutionary trajectories are likely to have multiple branches.} In
the absence of a complete theory, it is useful to see how the spectral
energy distributions vary over the parameter space
$(\mplan,\dotplan,\bplan,a)$. For the sake of definiteness, we fix the
planet radius so that $\rplan=10^{10}$ cm.

Figure \ref{fig:sedmass} shows how the spectral energy distributions
depend on the planet mass, which is varied over the range $\mplan$ =
0.1 -- 10 $M_J$. For all of the models, the mass accretion rate is
fixed at $\dotplan$ = $1M_J$/Myr, the semimajor axis $a$ = 5 AU, and
surface magnetic field strength $B_{\rm p}=500$ gauss. Two trends are
evident. First, as expected, the total luminosity of the planet/disk
system increases with mass, so that the larger objects are brighter.
Moreover, to leading order, $L\propto\mplan$. Second, the spectral
energy distributions become bluer with increasing planet mass. For
planets of higher mass, the planetary surface is hotter and the total
optical depth through the infalling envelope is lower, so that a
smaller fraction of the energy is emitted at longer wavelengths. 
These trends are consistent with previous numerical studies 
\citep{szulagyi2019}.

The dependence of the spectral energy distributions on the mass
accretion rate is shown in Figure \ref{fig:sedmdot}, where $\dotplan$
= 0.1 -- 10 $M_J$/Myr. In this case, the planet mass is fixed at
$\mplan=1M_J$, the semimajor axis $a$ = 5 AU, and surface magnetic
field strength $B_{\rm p}=500$ gauss. As before, two trends are
present (compare with Figure \ref{fig:sedmass}).  The total luminosity
of the planet/disk system increases with the mass accretion rate,
where the luminosity scales as $L\propto\dotplan$ to leading order. 
In this case, however, the the spectral energy distributions become
redder with increasing luminosity (due to the larger mass accretion
rate). The increase in $\dotplan$ corresponds to an increase in the
column density, so that more of the central source radiation is
absorbed by the infalling envelope and re-radiated into the infrared.

The effects of varying the magnetic field strength are shown in Figure
\ref{fig:sedbmag}. In this case, the $\bplan$ varies over the range 10
-- 1000 gauss.  The planet has mass $\mplan=1M_J$, accretion rate
$\dotplan=1M_J$/Myr, and semimajor axis $a$ = 5 AU. For the smallest
field value, $\bplan=10$ gauss, the magnetic truncation radius $R_X$
is smaller than the planetary radius, so that the disk extends all the
way to the planetary surface. For stronger magnetic fields with
$\bplan$ = 100 and 1000 gauss, the radius $R_X/\rplan$ = 1.5 and 5.6,
respectively, and magnetic truncation becomes important. The larger
values of $R_X$ lead to modest increases in the luminosity, due to
less rotational energy being stored in the circumplanetary disk.
Larger truncation radii $R_X$ also result in bluer planet/disk
spectra, so that the extinction (for fixed column density) is
greater. As a result, a double-humped form of the spectral energy
distribution is produced for systems with large field strengths
$\bplan$.

Figure \ref{fig:sedarad} shows how the spectral energy distributions
vary with the semimajor axis of the orbit of the forming planet.  Here
the planet mass $\mplan$ = 1 $M_J$, the accretion rate $\dotplan$ = 1
$M_J$/Myr, and the magnetic field strength is fixed at $\bplan=350$
gauss. Results are shown for semimajor axes in the range 0.1 -- 10 AU.
The spectral energy distributions become wider as the semimajor axis
increases. The most important effect is that the radius of the
circumplanetary disk scales as $\rdisk\propto{R_H}\propto{a}$. For 
the lower end of the orbital range ($a=0.1$ AU), the disk radius
becomes comparable to the magnetic truncation radius. In this regime,
most of the radiation from the central source is emitted from a single
surface temperature. Moreover, the column density is larger, so that
more of the central source photons are absorbed and re-emitted by the
envelope. The resulting spectral energy distribution thus approaches a
blackbody form. For larger semimajor axes, the column density is
smaller, and the spectral energy distributions are broader and bluer.

\section{Conclusion} 
\label{sec:conclude} 

This paper has constructed an analytic approach to the late stages of
accretion for the formation of gaseous giant planets. This work is
applicable to the third stage of the formation process when the planet
accretes the majority of its mass. 

\subsection{Summary of Results} 
\label{sec:summary} 

Our main results can be summarized as follows:

We have developed an infall solution to describe the density and
velocity fields of the material falling toward a forming giant
planet. The flow enters into the sphere of influence of the planet
near the Hill radius, and then smoothly approaches ballistic
trajectories.  In this approximation, the density distribution
$\rho(r,\mu)$ (given by equations [\ref{orbit}] and [\ref{density}])
and the velocity field ${\vec v}(r,\mu)$ (given by equations
[\ref{vzero} -- \ref{vphi}]) are axisymmetric. The material with the
highest specific angular momentum falls onto a circumplanetary disk
with radius given by the centrifugal barrier ($\rdisk=R_C=R_H/3$; 
equation [\ref{rcent}]).

To gain insight into the nature of the region surrounding the forming
planet, we have constructed the equivalent spherical density
distribution (equations [\ref{rhosphere}--\ref{asphericity}]). The
total column density $N_{\rm col}$ (given by equation [\ref{column}])
of the envelope plays an important role in determining the spectral
energy distribution of the forming body. The envelope surrounding the
planet is predicted to be marginally optically thick to the radiation
emitted from the central planet/disk systems, but is predicted to be
optically thin to radiation emitted by the envelope itself.

Most of the material falling toward the planet has too much specific
angular momentum to reach the planetary surface. Instead, material
falls onto a circumplanetary disk with radius $R_C$ determined by
conservation of angular momentum (equation [\ref{rcent}]) and an inner
radius $R_X$ determined by magnetic truncation (equation
[\ref{rmag}]). Given the expected disk radius and mass accretion rate
onto the disk, the time scale for viscous evolution is shorter than
that for disk mass accumulation provided that the dimensionless
parameter $\alpha\gta10^{-5}$. As result, the disk is expected to
reach a steady state, with low mass compared to the central planet,
and with a surface density distribution $\Sigma(r)$ given by equations
(\ref{sssigma}) and (\ref{gensigma}).

Magnetic fields play an important role in determining the geometry of
the inner disk and inner infall region. The balance between magnetic
pressure and mass accretion rate through the disk defines the
truncation radius $R_X$ (equation [\ref{rmag}]), which specifies the
inner disk edge. The field configuration also defines a capture radius
$R_\sigma$ (equation [\ref{rcap}]), which is determined by the balance
between the ram pressure of infalling gas and the magnetic pressure at
the point where the field lines are horizontal (see Figure
\ref{fig:capture}). 

The forming planet generates power from several contributions (see
Section \ref{sec:luminosity}). The relevant luminosity sources include
shocks from direct infall onto the planetary surface, accretion from
the inner disk edge onto the planet, viscous accretion through the
circumplanetary disk, as well as both shocks and mixing from material
falling onto the disk surface. The planet generates additional
internal power through gravitational contraction. The total luminosity
is comparable to the benchmark power scale given by equation
(\ref{lumzero}).

With the properties of the components specified, we have calculated
radiative signatures of the planet forming process. The central planet
and disk produce an intrinsic spectral energy distribution that
corresponds to surface temperatures in the range $T=100-1000$ K
(Figure \ref{fig:source}). The infalling envelope attenuates the
central source luminsity, which is re-radiated at longer wavelengths,
primarily in the infrared. The resulting spectral energy distributions
$\nu L_\nu$ are significantly broader than single-temperature
blackbodies and show systematic variations with planet mass $\mplan$,
accretion rate $\dotplan$, magnetic field strength $\bplan$, and
semimajor axis $a$ (these trends are illustrated in Figures
\ref{fig:emergent} -- \ref{fig:sedarad}).

\subsection{Star Formation versus Planet Formation} 
\label{sec:compare} 

The infall-collapse-disk solutions constructed here for the formation
of giant planets are analogous to those found earlier for the star
formation problem (see the review of \citealt{shu1987}). It is useful
to outline the ways in which the formation processes are similar ---
and different --- for the two cases. In both settings, angular
momentum plays a key role in determining the geometry of the incoming
material. Both star formation and planet formation produce nearly
pressure-free collapse flow, with high angular momentum orbits that
lead to the formation of a disk structure. Significantly, most of the
mass initially falls onto the disk in both cases. The resulting
dynamic range, as determined by the ratio of the disk radius to that
of the central body, is much larger for star formation ($R_C/R_\ast
\sim 10^4$) than for planet formation ($R_C/\rplan\sim100$). Stars
form much faster, with typical formation times of order 0.1 Myr,
compared to 1 Myr for planets.\footnote{For both star and planet
  formation, these time scales have a range, and can be somewhat
  larger than the values quoted here. Nonetheless, the time scale for
  star formation is about an order of magnitude shorter than that of
  planet formation.}  With a longer formation times and smaller
dynamic ranges, circumplanetary disks require much smaller viscosity
to reach a steady-state where accretion through the disk keeps pace
with infall onto the disk. Although the source of viscosity remains
under study, modest values of the parameter $\alpha\sim10^{-4}$ are
sufficient to keep circumplanetary disks in steady-state (and such
values are expected to be realized). Much larger viscosity levels are
required for star-forming disks to reach steady-state. As a result,
circumstellar disks are expected to experience gravitational
instability during their early formative stages. In addition, the
circumstellar disks are subject to episodic accretion, which produces
FU Orionis outbursts. It remains to be seen if circumplanetary disks
are subject to the same episodic behavior.

The dynamic range in temperature is also much larger for star
formation, where the stellar surface has typical temperature
$T_\ast\sim6000$ K, and the outer part of the infall region has
interstellar temperatures $T_{\rm ism}\sim10-30$ K. For planet
formation, the planetary surface temperature is significantly smaller
$\tplan\sim$ $1000-1500$ K, and the temperature at the outer boundary
is typically $T_H\sim100$ K (for $a\sim5$ AU). The temperature range
for star formation (factor of $\sim300$) is more than an order of
magnitude greater than the range for planet formation (factor of
$\sim12$). Planets forming at smaller semimajor axes (e.g., $a\sim0.1$
AU), have even smaller temperature ranges.

In both star and planet formation, the rate ${\dot M}$ at which the
growing body gains mass is of fundamental importance. In the case of
star formation, ${\dot M}_{\rm core}\approx \vsound^3/G$ is determined
by the pre-collapse conditions in the background molecular cloud. This
quantity determines the rate at which the entire star/disk system
gains mass from its parental cloud. Dissipative processes within the
circumstellar disk then determine the smaller rate ${\dot M}_{\ast{\rm
    d}}$ at which the disk accretes material onto the star.  In the
context of planet formation, the background circumstellar disk plays a
role similar to the cloud in star formation.  In the latter case,
however, the semimajor axis of the planet determines the Hill radius,
which in turn affects the accretion rate $\dotplan$ into the sphere of
influence (where $\dotplan$ is a fraction of ${\dot M}_{\ast {\rm
    d}}$). Unlike stars, forming planets are thus subject to a
systematic variation in background conditions, with corresponding
variations in ${\dot M}$.

\subsection{Discussion}
\label{sec:discuss} 

To help organize our understanding of the late stages of the 
planet formation process, we can consider the ordering of the 
relevant length scales:
\be
\rplan < R_X \sim R_\sigma \ll R_C \sim \rdisk < R_H \sim H \ll a \,.
\label{order} 
\ee 
During the phase when the growing planet accretes most of its gas, the
protoplanet thus develops a hierarchical structure.  Most of the mass
entering the sphere of influence from the background cirumstellar disk
falls onto the circumplanetary disk, which is much larger than the
planet itself. Magnetic fields lead to a truncation of the inner disk
and provide a larger effective capture cross section at a scale of
several planetary radii. The nominal disk radius is one third of the
Hill radius, which is roughly comparable to the scale height of the
circumstellar disk. For the formation of giant planets near or beyond
the ice-line ($a\gta3$ AU), all of these scales are much smaller than
the semimajor axis at the location of the planet. For hot Jupiters
($a\lta0.1$ AU), however, the length scales that define the outer disk
($\rdisk$, $R_C$, $R_H$, $H$) become comparable to the magnetic radii
($R_X$, $R_\sigma$). This reordering of length scales implies that
giant planet formation could proceed differently in the inner and outer
(circumstellar) disk, with a boundary between the two regimes at
roughly $a\sim0.1$ AU. 

The relevant time scales also have a well defined ordering, 
\be
t_{\rm disk} < t_{\rm orb} \ll t_{\rm vis} \ll t_{\rm form} 
\lta t_{\rm kh} \,. 
\ee
The orbital time scale of the forming planet (around the star) is 9
times longer than the orbital time scale at the outer edge of the
circumplanetary disk (around the planet), so that $t_{\rm orb}$ =
$9t_{\rm disk}$.  The viscous evolution time scale $t_{\rm vis}$ =
$R_C^2/\nu$ of the circumplanetary disk is much longer than the orbit
time, but much shorter than the formation time scale $t_{\rm form}$ =
$\mplan/\dotplan$. As a result, the disk can evolve to a steady-state
configuration (see equation [\ref{sssigma}]). The Kelvin-Helmholtz
time scale $t_{\rm kh}$ sets the times scale for the evolution of the
internal planetary structure, and is comparable to the formation time
if the luminosity is determined by the scale $L_0$ from equation 
(\ref{lumzero}), but somewhat longer if we use only the internally
generated luminosity to define $t_{\rm kh}$.

In general, the Hill radius $R_H$ provides the outer boundary for the
planetary sphere of influence, including in this present work.
Although this ansatz provides a good first approximation, the picture
is more nuanced. Numerical simulations indicate that some of the
material that initially flows into the Hill sphere promptly flows back
out (e.g., \citealt{lambrechts2019}).  The treatment of this paper
accounts for this loss by taking the mass infall rate $\dotplan$ to
include only the material that initially stays within the Hill sphere
and falls toward the planet. A related complication is that the Hill
radius is only an approximation to the true boundary. Pressure and
magnetic corrections lead to smaller values for the effective boundary
between the planet and its background disk (Appendix
\ref{sec:hillpressure}). A smaller boundary radius, in turn, leads to
a smaller centrifugal barrier $R_C$, which sets the nominal radius of
the circumplanetary disk (equations [\ref{orbit}--\ref{rcent}]). 
Two dimensional simulations find that the expected disk radius  
at $R_C=R_H/3$ provides a good approximation \citep{martin2011},
whereas three dimensional simulations find somewhat smaller values for
$R_C$ \citep{fung2019}. Three dimensional simulations (e.g., 
\citealt{lambrechts2019}) show that infall takes place preferentially
along the poles of the system. This confinement results in a smaller
centrifugal radius ($R_C\to R_C[1-\mu_b^2]$) for a given planetary
mass and a somewhat steeper surface density distribution.  Finally,
some of the material that initially falls onto the circumplanetary
disk must eventually flow back out into the circumstellar disk. 
Viscous torques cause the circumplanetary disk to spread, so that most
of the mass flows inward, but some mass flows outward to conserve
angular momentum. As a result, some fraction of the disk material must
diffuse beyond the outer boundary (given by the generalized Hill
radius) and rejoin the background circumstellar disk.\footnote{For
completeness, we note that if the disk supports outward flow
(decretion) over an extended range of radii, then the prospects for
moon formation are enhanced \citep{batmorby2020}. Since accretion
(inward flow) must take place in order for the planet to gain
significant mass (Section \ref{sec:noaccretion}), such decretion flow
must take place during late evolutionary stages, after the planet has
acquired most of its material.} The analytic treatment developed in
this paper is robust, in that it provides solutions for any value of
the centrifugal radius. On the other hand, additional work is required
to determine the expected values of $R_C$ to higher accuracy.

This analytic approach highlights how planet formation can vary with
the location of the planetary orbit. For a given planetary mass, the
physical size of the Hill radius and the circumplanetary disk radius
grow with increasing semimajor axis $a$. Disk viscosity must play an
increasingly important role for more distant planets. This trend has
the effect of making planet formation (through the core accretion
paradigm) more difficult with increasing $a$. This trend acts in
addition to the previously discussed effect that the formation of
planetary cores takes longer for larger orbits. In the opposite limit,
for small orbits akin to those of hot Jupiters, accretion through
circumplanetary disks is even more efficient. However, for $a\lta0.1$
AU, the expected magnetic truncation radius becomes smaller than the
disk radius, and infall onto the entire planet/disk system can be
suppressed. 

This paper also provides an analytic description of the radiative
signature of forming gas giants. Observations of dust emission from
circumplanetary disks (which are embedded within circumstellar disks)
have just become technologically possible \citep{benisty2021} and more
such detections are expected.  Current observations of circumplanetary
disks are roughly consistent with theoretical expectations ---
including those of this paper. As future observations further
elucidate the properties of forming planets, the analytic treatment of
this paper can be readily adopted and generalized to provide
theoretical descriptions of the formation process.

\medskip 
{\sl Acknowledgments:} 
We are grateful to A. Adams, M. Meyer, and D. Stevenson for useful
discussions. We thank an anonymous referee for many helpful comments.
This work was supported by the University of Michigan, Caltech, the
Leinweber Center for Theoretical Physics, and by the David the Lucile
Packard Foundation.

\appendix
\section{Approximation Control} 
\label{sec:control} 

The main text presents a working analytic model for the infall of
material onto a forming planet and its circumplanetary disk, along
with solutions for the evolution of the disk. With the model in place,
this Appendix outlines the requirements for the solutions to be self
consistent. More specifically, we estimate the magnitude of the errors
introduced in making the approximations necessary for an analytic
treatment. 

The infall collapse solution is constructed using a constant 
mass $\mplan$ for the planet. Consistency requires that the time 
scale for the planet mass to change must be much longer than the 
time required for parcels of gas to fall inward. The evolution time 
for the mass of the planet is expected to be 
$t_{\rm acc}=\mplan/\dotplan\sim$ 1 Myr. For comparison, 
the infall time from the outer boundary is given by 
\be
t_{\rm in} \sim \left({G\mplan\over R_H^3}\right)^{-1/2} = 
\left({3GM_\ast\over a^3}\right)^{-1/2} \approx 1\,{\rm yr}\,
\left({M_\ast\over1M_\odot}\right)^{-1/2} 
\left({a\over5\,{\rm AU}}\right)^{3/2} \,.
\ee
We thus find $t_{\rm in}\ll t_{\rm acc}$ as required. Planets can also change
their radial locations (migrate) during the time when they acquire
mass. The typical migration time scales are of order $\sim1$ Myr, so
that the infall time scale is much shorter, and the analytic solutions
of this work are applicable.

Another assumption is that the infalling gas follows ballistic
trajectories, which require that the pressure forces are much smaller
than the gravitational forces. At the outer boundary, given by the
Hill radius, this requirement can be written in the approximate form
\be 
{G \mplan \over R_H^2} \gg {1\over\rho} {dP\over dr} \approx 
\Omega^2 R_H G(\theta) \,,
\ee 
where the second equality follows from the assumption that the
circumstellar disk is locally in hydrostatic equilibrium. The
geometrical factor $G(\theta)$ is unity in the $\hat z$ direction,
which we take to be along the pole of the planet, whereas $G\sim H/a$
along the equatorial directions. Using $\Omega^2$ = $GM_\ast/a^3$ and
the definition of the Hill radius, we find that the left-hand side of
the equation is larger than the right-hand side by a factor of $3$ for
polar directions and a factor of $\sim30$ for equatorial
directions. Alternately, we can compare the depth of the potential
well $G\mplan/R_H$ with the local sound speed $v_s^2$. For a Jovian
planet forming at $a=5$ AU, the gravitational potential well at $R_H$
corresponds to a speed $\sim1.6$ km/s.  The sound speed is given by
$\vsound\approx0.72$ km/s ($T$/130K)$^{1/2}$, so the pressure is
subdominant by a factor of $\sim5$ for the values used here.
Moreover, pressure forces will become less important as the flow moves
inward, provided that the equation of state is not too stiff.
Specifically, studies of collapse solutions for star formation show
that if the dynamic equation of state\footnote{The dynamic equation of
  state describes the thermodynamics of the gas as in flows inward and
  becomes compressed. It remains possible for the equation of state
  that determines the pre-collapse state to have a different form,
  where the latter is sometimes called the static equation of state.}
has the polytropic form $P\propto\rho^\gamma$, the requirement
$\gamma<5/3$ allows for trajectories to remain ballistic
(\citealt{fatuzzo2004}; see also \citealt{mckee1999}). On the other
hand, a sufficiently stiff equation of state (corresponding to
inefficient cooling and hence large $\gamma$) can result in
non-negligible pressure terms. Numerical simulations 
\citep{fung2019} show similar trends, where an isothermal equation of
state leads to the formation of rotationally supported disks, but
adiabatic simulations result in extended structures. 

The orbit equation that describes infalling trajectories was derived
for the gravitational potential of a point mass. However, the
extended structure of the circumplanetary disk will cause orbits
to precess relative to this approximation. The first non-zero 
correction to the potential corresponds to the quadrupole term, 
which decreases rapidly with radial distance. In addition, the 
disk mass is expected to be small, with $\mdisk/\mplan\sim$
${\cal R}\sim0.01$. 

The infalling trajectories were taken to be zero energy orbits. Since
the incoming gas enters the sphere of influence at the Hill sphere,
the total speed of the gas must be given by $v^2=2G\mplan/R_H$ for
consistency. Here we assume that the specific angular momentum is
given by $j_\infty=\Omega R_H^2$, so that one component of the
velocity is $\Omega R_H$. Since $(\Omega R_H)^2 = G\mplan/3R_H$, we
are implicitly assuming that the other incoming velocity components
add up to $v_\perp^2=5G\mplan/3R_H$. If the actual (non-rotational)
velocity components were zero, then the zero-energy approximation
corresponds to a relative error in energy of $5r/6R_H$. This energy
inconsistency is only $\sim0.16\%$ for orbits that land on the
planetary surface, but grows to $\sim28\%$ for orbits that fall to the
outer edge of the disk (see also \citealt{mendoza2009} for further 
generalizations of the infall trajectories). 

The initial conditions for infall, and hence the resulting solutions,
are assumed to be azimuthally symmetric. Numerical simulations of the
planet formation process indicate that the incoming flow can be
concentrated into streamers, thus breaking the symmetry.  However,
this complication has only a modest affect on the properties of the
circumplanetary disk and the forming planet. The incoming material
primarily falls onto the disk, where it enters a nearly Keplerian
orbit around the planet. The orbital time scale at the outer disk edge
is much shorter ($\sim1$ year) than the time scale on which the planet
gains mass ($\sim1$ Myr). As a result, differential rotation rapidly
spreads material over the orbit so that the disk becomes azimuthally
symmetric. On the other hand, for purposes of calculating the spectral
signatures (Section \ref{sec:radiation}), we carry out the radiative
transfer for a fully spherical envelope. These spherical solutions
would be modified with non-axisymmetric infalling envelopes. The
results are model dependent, however, and this generalization is
left for future work.

The treatement has ignored magnetic field effects in determining the
properties of the infalling envelope. In order for magnetic fields to
influence infall onto the disk to a significant degree, the magnetic
pressure must compete with the ram pressure of the infall at the 
outer disk edge. This consideration implies the consistency condition 
\be
{B_C^2 \over 8\pi} \lta \rho v^2  = {\dotplan v_C \over 4\pi R_C^2} \,,
\ee
where the subscripts indicate that the quantities are evaluated at the
disk edge. The magnetic field strength $B_C$ at the edge is less than
or equal to the value indicated by flux-freezing, i.e., $B_C \le B_0$
$(R_H^2/R_C^2)=9B_0$, where $B_0$ is the magnetic field strength of
the circumstellar disk material evaluated outside the Hill sphere
(note that $B_0$ is expected to be a function of semimajor axis
$a$). We obtain a sufficient condition on $B_0$ by combining the above
expressions to obtain 
\be
B_0 \lta \sqrt{2\over9} 
\left[ {G \mplan \dotplan^2 \over R_C^5} \right]^{1/4} 
\approx 36 \, {\rm mG} \, 
\left({\mplan\over M_J}\right)^{-1/6} 
\left({\dotplan\over1M_J/{\rm Myr}}\right)^{1/2} 
\left({a \over 5\,{\rm AU}}\right)^{-5/4} \,.
\ee
Since most estimates for disk magnetic field strengths are below this
value, the disk can form as described in the main text.  Note that if
flux-freezing holds down to smaller radii, $r\ll R_C$, then the inner
regions could be magnetically affected for sufficiently large values
of $B_0$. In this case, magnetic fields could cause the infalling
parcels of gas to fall to larger radii within the disk, thereby
leading to smaller shock luminosity from the disk surface. As long as
the disk viscosity is large enough, however, the disk would accrete as
before and deliver the same amount of material to the planet. Moreover, 
if the volume density of the gas exceeds a threshold of $n\sim10^{11}$
cm$^{-3}$, coupling between the magnetic field and the fluid is 
compromised, and the magnetic flux can be dynamically redistributed
\citep{nishi1991,desch2001}.

Another consideration for magnetic field effects is the time required
for the resistivity $\eta$ to change the field structure. The
resistivity acts as a diffusion constant, so that the time scale for
field evolution over a length scale $\ell$ is given approximately by 
$t \sim \ell^2/\eta$. If we require that the resistivity is large 
enough so that this time scale is less than the free-fall time, 
the constraint on $\eta$ at the Hill radius $R_H$ takes the form 
\be
\eta \gta \sqrt{G\mplan a} 
\left({\mplan\over3M_\ast}\right)^{1/6} = 8.3 \times 10^{17} 
{\rm cm}^2 {\rm s}^{-1} \left({a \over5{\rm AU}}\right)^{1/2} 
\left({\mplan\over m_J}\right)^{2/3}\,. 
\ee
Estimates for the appropriate resistivity in circumstellar disks, 
which provide the background environment for the collapse flow, 
vary widely. Estimates for the resistivity include the range 
$\eta=10^{16}-10^{19}$ cm$^2$ s$^{-1}$ \citep{stepinski1992}, 
$\eta\approx10^{16}$ cm$^2$ s$^{-1}$ \citep{wardle1999}, 
$\eta\approx10^{19}$ cm$^2$ s$^{-1}$ \citep{desch2001}, and 
$\eta\approx10^{20}$ cm$^2$ s$^{-1}$ \citep{shu2006}. In addition 
to the electrical resistivity, ambipolar diffusion can provide a
substantial contribution $\eta_{\scriptstyle AD}$ 
$\sim B^2/(4\pi\gamma\rho_i\rho_n)$, where $\gamma$ is the ion-neutral
drag coefficient and $\rho_i$ and $\rho_n$ are the ion and neutral
densities (see also \citealt{nakano2002}). In summary, most estimates
indicate that $\eta$ is large enough for magnetic fields to diffuse
outward fast enough for the flow to remain in the kinematic regime,
although further work must be carried out.

\section{Planetary Magnetic Field Strength} 
\label{sec:magfield} 

Since the magnetic field can play an important role in accretion of
material onto a forming planet, we need to estimate the expected field
strength.  One now-standard scaling law assumes that the field
strength is proportional to the energy generation rate due to buoyancy
forces (e.g., \citealt{christensen2009,yadav2017}). 
This law can be written in the form 
\be
{B^2 \over \rho} \sim \ell^{2/3} \power^{2/3}\,,
\ee
where $\power$ is the energy production rate per unit mass and $\ell$
is a length scale associated with the size of a convective cell. We
expect the length $\ell$ to be a fraction of the planetary radius 
\be
\ell = f_\ell \rplan\,. 
\ee
In this expression, $B$ is the field strength in the dynamo region. 
The surface field will be a fraction of this value so that
\be
\bplan = f_B B \, . 
\ee
Finally, we expect the power transported through the planetary
interior to be a fraction $f_P$ of the total power generated by the
object, where the latter is dominated by accretion. As a result, we
can write 
\be
\power = f_P {G \dotplan \over \rplan} \,.
\ee
Significantly, we expect the three factors $(f_\ell,f_B,f_P)$ 
to all be less than unity. The surface field then takes the form 
\be
\bplan^2 = f_T {3\mplan f_\rho\over4\pi\rplan^3} 
\left( G \dotplan \right)^{2/3} 
\qquad {\rm where} \qquad 
f_T \equiv f_B^2 f_\ell^{2/3} f_P^{2/3} \,.
\label{bvmass} 
\ee
The factor $f_\rho$ takes into account the difference between the mean
density of the planet and that of the convective cells; the second
equality collects the other dimensionless factors into a single
quantity $f_T<1$. Inserting typical values, we find the field 
strength estimate 
\be
\bplan \approx 1600\,{\rm gauss}\,\,\,f_T^{1/2} 
\left({\rho \over 1\,{\rm g}\,{\rm cm}^{-3}}\right)^{1/2} 
\left({\dotplan \over 1\,M_J\,{\rm Myr}^{-1}}\right)^{1/3} \,.
\ee
The factor $f_T$ can be much smaller than unity, so that the numerical
value provides an appoximate upper limit to the expected field strength.
Nonetheless, substantial magnetic fields can be produced within
forming planets if sufficient accretion energy is converted into
convective motions. For example, if we take $f_B=1/2=f_\ell$ and 
$f_P=1/10$, then $\bplan\approx300$ gauss. 

\section{Corrections to the Hill Radius} 
\label{sec:hillpressure} 

In the treatment of this paper, the Hill radius $R_H$ marks the
boundary between the circumstellar disk and the region where the
gravitational influence of the planet dominates. The Hill radius sets
the centrifugal radius for the collapse flow (where $R_C=R_H/3$) and
thus determines the disk radius ($\rdisk=R_C$). Given that this 
boundary plays an important role in the planet formation problem, 
this Appendix considers possible corrections. 

The effects of pressure can be accounted for in the derivation of the
Hill radius. The net effect of including pressure is to reduce the
effective mass of the planet for purposes of defining the extent of
its influence. The pressure-corrected Hill radius can thus be 
written in the form 
\be
R_{H*} = a \left({\fpc \mplan \over 3M_\ast}\right)^{1/3} \,,
\ee
where $\fpc$ is a dimensionless factor less than unity. Using this
result in the definition of the centrifugal radius (\ref{rcent}), 
we find the pressure-corrected form 
\be
R_{C*} = {\fpc \over 3} R_{H*} = \fpc^{4/3} {R_H \over 3} \,. 
\ee
We expect $\fpc\sim2/3$ (see Appendix \ref{sec:control}) so that 
the correction factor is $\fpc^{4/3}\sim0.58$ and the disk radius
$\rdisk$ = $R_{C*} \sim R_H/5$ (where $R_H$ is the usual, 
uncorrected, Hill radius). 

With this correction, the mass scale $M_0$ that defines when
the centrifugal radius exceeds the planetary radius becomes 
\be
M_0 = \left({3\over \fpc}\right)^4 \left({\rplan\over a}\right)^3\,.
\ee
Equation (\ref{pfraction}) gives the planet mass $\mplan$ for a given
total mass $M$ that has fallen into the Hill sphere. The planet mass
$\mplan \propto M_0^{1/3} \propto \fpc^{-4/3}$. As a result, larger
pressure suppression leads to a smaller centrifugal barrier and more
mass being directly accreted onto the planet.

The discussion thus far uses the Hill radius as the boundary between
the forming planet and the background circumstellar disk. An alternate
choice is to the use the sphere of influence from astrodynamics (e.g.,
\citealt{danby2003}). The standard expression for this boundary has 
the form
\be
R_S = {a \over 2^{1/5}} 
\left({\mplan\over M_\ast}\right)^{2/5} \,.
\ee
For planets with Jovian mass, $\mplan=1M_J$, we find that
$R_S\approx0.79R_H$, so the difference is modest. However, the
centrifugal barrier $R_C \propto R_S^4$, so that the correction can be
substantial. In addition, the scaling with mass is somewhat
different. Moreover, using the sphere of influence $R_S$ allows for a
staightforward way to include pressure forces into the definition of
the boundary.  As outlined in Appendix \ref{sec:control},
gravitational forces due to the planet are (at least) 3 times greater
than the pressure forces at the Hill sphere.  This finding implies
that the gravitational acceleration used in the derivation of the
sphere of influence is reduced by a factor of $f\sim2/3$.  The
pressure-corrected sphere of influence can thus be written in 
the form 
\be
R_S = {a \over 2^{1/5}} 
\left({f \mplan\over M_\ast}\right)^{2/5} \,.
\ee
This radius thus defines the specific angular momentum at the 
starting point for collapse. The corresponding centrifugal 
radius becomes 
\be
R_C = a \left({f^2 \over 2}\right)^{4/5} 
\left({\mplan\over M_\ast}\right)^{3/5} \,.
\ee
Note that both the coefficient and the scaling with mass 
are different than in the previous case (using the Hill radius).

\section{Disk Solution Beyond the Centrifugal Barrier} 
\label{sec:diskextend} 

The steady-state solution found in the main text applies over the
radial range $\rplan\le{r}\le R_C$. However, the disk will spread
beyond the centrifugal barrier. In this regime, the steady-state
solution for the surface density is given by 
\be
r^{1/2} {d \over dr} \left( r^{1/2} \nu \Sigma 
\right) = constant\,,
\ee
which has a solution of the form 
\be
\nu \Sigma = C_0 + C_1 u^{-1/2} \,,
\ee 
where $u=r/R_C$ (and $u>1$ in this regime). The constants $C_0$ and
$C_1$ can be specified by requiring continuity of both $\Sigma(u)$ and
its spatial derivative across the boundary $u=1$ ($r=R_C$). These
matching conditions produce the constraints 
\be
C_0 + C_1 = {\dotplan \over 12} \qquad {\rm and} \qquad 
{C_1 \over C_0 + C_1} = 1 \,.
\ee
The constants thus have values $C_0=0$ and $C_1=\dotplan/12$.  
The disk surface density thus decreases more rapidly (by a factor 
of $r^{-1/2}$) beyond the centrifugual barrier, which defines the
nomimal disk radius.

Notice also that the disk must spread outward in order to satisfy
conservation of angular momentum globally. Before collapse, the total
angular momentum within the Hill sphere for a planetary mass $\mplan$
is given by 
\be
J = {2\over5} \mplan R_H^2 \Omega \,. 
\ee
After collapse, the planet and its disk must account for this amount
of angular momentum.  If we consider the angular momentum of the
spinning planet to be negligible in comparison, then some mass
increment $\Delta{M}$ must orbit a large distance $R$ in order to
carry the angular momentum $J$. Setting $R=R_H$ and solving for the
required mass increment, we find 
\be
{\Delta M \over \mplan} = {2 \over 5 \sqrt{3}} 
\approx 0.23\,.
\ee
In other words, roughly one quarter of the mass must spread 
out to the Hill sphere so that the material forming the planet 
can lose enough angular momentum to reach small radial distances. 
This results holds in the absence of any redistribution of 
angular momentum before the collapse. The action of pressure 
and magnetic fields can reduce the effective centrifugal barrier 
(see Appendix \ref{sec:hillpressure}), leading to a lower estimate 
for the mass increment $\Delta M$.

\section{Evolution} 
\label{sec:evolution} 

The analytic model developed in this paper provides solutions for the
properties of the planet/disk/envelope system for any combination of
the input parameters. These output properties include the envelope
density distribution $\rho(r,\mu)$, velocity field ${\vec v}(r,\mu)$,
column density $N_{\rm col}$, disk radius $\rdisk=R_C$, magnetic
truncation radius $R_X$, magnetic capture radius $R_\sigma$, and the
resulting spectral energy distributions $\nu L_\nu(\nu)$. The required
input parameters include the planet mass $\mplan$, radius $\rplan$,
accretion rate $\dotplan$, magnetic field strength $\bplan$, and the
semimajor axis $a$ of the forming object. In this treatment, we
consider the planet radius as a slowly varying parameter and fix its
value ($\rplan\sim10^{10}$ cm). However, one could consider the case
where cooling is inefficient and the planetary radius is much larger. 
In the present context, we are left with the input variables 
$(\mplan,\dotplan,\bplan,a)$ that specify the system.

In a complete theory, which is not yet available, the parameters
$(\mplan,\dotplan,\bplan,a)$ are a function of time. For example, one
expects $\mplan\sim\int\dotplan dt$.  Working within this analytic
model, we can consider evolutionary scenarios in which the parameters
are interdependent. This dependence reduces the number of parameters
and allows for the construction of an evolutionary sequence. 

As one example, the mass accretion rate is likely to depend on planet
mass. Many authors have suggested that the mass accretion rate
$\dotplan$ is a power-law function of the Hill radius so that 
\be
\dotplan = \dotzero \left({R_H \over R_0}\right)^p = 
\dotzero \left({\mplan \over M_0}\right)^{p/3} \,, 
\label{mdothill} 
\ee
where $p=4$ is the most common choice. If we specify the mass scale
$M_0=M_J$, then the relevant input parameter is the accretion scale
$\dotzero$. We expect $\dotzero\sim1M_J$/Myr so that giant planets
can gain enough mass during the lifetimes of circumstellar disks
(where $\tau\sim3$ Myr; \citealt{jesus}). In addition, one can show
that basic form of the planetary mass function can be reproduced under
the assumption that accretion ends due to the removal of gas from the
circumstellar disk \citep{adams2021}. In any case, if the mass
accretion rate is determined by equation (\ref{mdothill}), the number
of input variables is reduced.

The magnetic field strength $\bplan$ depends on both the planet mass
and the accretion rate. Although dynamo theory remains incomplete, we
can consider approximate scaling laws (Appendix \ref{sec:magfield}).
Using equation (\ref{mdothill}) to specify the mass dependence of the
accretion rate, in conjunction with equation (\ref{bvmass}), the field
strength scales as $\bplan \propto \mplan^{17/18}$. As a result, the
magnetic truncation radius from equation (\ref{rmag}) scales with mass
according to $R_X \propto \mplan^{1/63}$. The magnetic truncation
radius is thus nearly constant as the planet mass grows (in this 
scenario). 

For a given formation location, as set by the semimajor axis $a$, the
above considerations define an evolutionary scenario. For a given
time, equivalently planet mass, the accretion rate is specified by
equation (\ref{mdothill}) and the magnetic truncation radius is
approximately constant.  

Finally, we note that this formalism can also be used to consider
migration scenarios. If the planet moves inward according to
\be
{{\dot a} \over a} = - {1 \over \tau_a} \,,
\ee
then $(da/dM)/a = - 1/(\dotplan\tau_a)$. One can thus determine 
the planet location (semimajor axis $a$) as a function of planet 
mass. As long as the time scale for migration $a/{\dot a}$ 
sufficiently long (compared to the time required for the infall 
solution to adjust), the results of this paper continue to hold.

\end{document}